\documentclass[prb,superscriptaddress,twocolumn,eqsecnum,byrevtex]{revtex4}
\usepackage{multirow}
\usepackage{amsfonts}
\usepackage{amssymb}
\usepackage{amsmath}
\usepackage{color}
\usepackage{graphicx}
\usepackage{subcaption}
\usepackage{time}
\usepackage[colorinlistoftodos]{todonotes}
\usepackage{sidecap}

\begin{document}
%
%
\title{Two electrons in harmonic confinement coupled to light in a cavity}

\author{Chenhang Huang}
\affiliation{Department of Physics and Astronomy, Vanderbilt University, Nashville, Tennessee, 37235, USA}

\author{Alexander Ahrens}
\affiliation{Department of Physics and Astronomy, Vanderbilt University, Nashville, Tennessee, 37235, USA}

\author{Matthew Beutel}
\affiliation{Department of Physics and Astronomy, Vanderbilt University, Nashville, Tennessee, 37235, USA}
\author{K\'alm\'an Varga}
\email{kalman.varga@vanderbilt.edu}
\affiliation{Department of Physics and Astronomy, Vanderbilt University, Nashville, Tennessee, 37235, USA}

\begin{abstract}
The energy and wave function of a harmonically confined two-electron system coupled to light is calculated by separating the wave functions of the relative and center of mass (CM) motions. The relative motion wave function has a known quasi-analytical solution. The light only couples to the CM variable and the coupled equation can be solved with diagonalization without approximations. The approach works for any coupling strength. Examples of wave functions of light-matter hybrid states are presented.
\end{abstract}

\maketitle

\section{Introduction}
Analytically or numerically easily solvable systems (e.g. by "exact diagonalization") have always been important test grounds for models and approximations. Recently, there is an intense interest in strongly coupled light-matter systems \cite{doi:10.1021/acsphotonics.9b00648,Schafer4883,Ruggenthaler2018,Flick15285,Flick3026,Rokaj_2018,PhysRevLett.122.193603,PhysRevLett.121.113002,doi:10.1063/5.0012723,PhysRevB.98.235123,PhysRevLett.119.136001,Mandal,Cederbaum2021,doi:10.1021/acs.jpclett.8b02609,PhysRevLett.126.153603}.
In these systems, the light-matter coupling cannot be treated perturbatively. The electronic excitations and the photons are superimposed, forming hybrid light-matter excitations. In this regime, there are only a few analytical approaches available to test and develop efficient numerical methods. Reviews of the recent theoretical and experimental development can be found in Refs.\cite{Rivera2020,https://doi.org/10.1002/qute.201900140,Garcia-Vidaleabd0336}.

In this paper, we consider a two-electron system interacting via the Coulomb interaction, confined by a harmonic oscillator interaction  coupled to light in a cavity. The system is described on the level of the Pauli-Fierz (PF) nonrelativistic QED Hamiltonian. The two-electron 
system in harmonic oscillator confinement is a quasi-exactly solvable (QES) problem. The wave function can be written as a product of the wave functions of the relative and CM motion. The relative motion wave function can be expanded into infinite series. For certain oscillator parameters, this infinite series can be reduced to a recursion \cite{taut}. The wave function of the CM motion is a simple harmonic oscillator eigenfunction. We will show that the photons only
couple to the CM coordinate and the coupled CM
photon system can be solved by exact diagonalization. 

The two-particle systems have long been investigated due to their analytic and quasi-exact solvability, which provides straightforward intuition for the physical system under scrutiny as well as an excellent benchmark test for numerical computations. Examples of QES quantum systems are the two-dimensional (2D) harmoniums\cite{taut,taut2} and the hydrogen-like atoms in homogeneous magnetic fields\cite{taut3}. These QES problems have been generalized to relativistic cases as well \cite{karw2,vill}. For harmonium systems, the separability condition guarantees the quasi-exact solvability for the Schr\"odinger equation\cite{karw}, and linearly coupled oscillators have been studied under this condition\cite{turb}. For hydrogen-like models, solutions have been found for particular forms of the inhomogeneous magnetic fields  \cite{liu,down}. Examples of other known QES models include the planar Dirac electron in hydrogen-like atoms\cite{choo,choo2}, one-body problems in power-law central potentials\cite{agbo,karw3}, relativistic 2D pion in constant magnetic fields\cite{akhm}, and 1D and 3D regularized Calogero models\cite{pont,down2}. QES models with different forms of
confinements, e.g. two electrons in one\cite{loos2} or two\cite{guo} 1D rings, two electrons on the surface of the n-sphere (spherium)\cite{loos,loos3}, have also been studied. 

The exact or even the numerical solution for light-matter coupled systems is very difficult even on the level of a minimal coupling Hamiltonian in the long-wavelength limit \cite{QEDHAM}, because the photons substantially increase the number degrees of freedom of the system. Theoretical approaches have been developed to tame the light-matter coupled systems using approximations and transformations \cite{FriskKockum2019,PhysRevA.98.043801,doi:10.1021/acsphotonics.7b01279,PhysRevLett.126.153603,doi:10.1021/acs.jpclett.0c01556,doi:10.1021/acs.jpclett.0c01556,QEDHAM,PhysRevLett.123.083201,
PhysRevLett.126.153603,PhysRevB.100.121109,10.21468/SciPostPhys.9.5.066,PhysRevResearch.3.023079,doi:10.1021/acs.jpclett.0c01556,rokaj2021free}. In Refs. \cite{PhysRevA.98.043801,doi:10.1021/acsphotonics.7b01279}, an electron in a 2D potential coupled to a single photon mode is used as a numerical benchmark test. The spatial part of the wave function is represented on a real space grid and coupled to the Fock space of the photons. The Hamiltonian of the system can be diagonalized in this representation and the light coupled wave function can be studied. In Ref. \cite{doi:10.1021/acs.jpclett.0c01556}, the spatial wave function of the He, HD$^+$, and H$_2^+$ three-particle system is represented using a 3D product of pseudospectral basis functions, and a few Fock spaces states of a single photon mode are coupled to the
spatial part. The energy and wave function is calculated by exact diagonalization of the PF Hamiltonian and the Jaynes–Cummings limit for electronic and ro-vibrational transitions are studied. One-dimensional model systems of atoms and molecules \cite{doi:10.1021/acsphotonics.9b00648,PhysRevLett.123.083201} often using the Shin-Metiu potential \cite{doi:10.1063/1.468795} are also useful to describe potential energy surfaces in cavities and test numerical approaches. 

The free electron gas also allows analytical treatment \cite{rokaj2021free}. In Ref. \cite{rokaj2021free},  the free electron gas in cavity is analytically solved in the long-wavelength limit for an arbitrary number of non-interacting electrons. It is
found that the electron-photon ground state is a Fermi liquid containing virtual photons.

Approaches to reformulating the problem have also been proposed.  In Ref. \cite{PhysRevLett.126.153603}, the light and matter degrees
of freedom are decoupled using a unitary transformation. In the transformed frame, both the light and the matter Hilbert spaces can be truncated systematically to facilitate an efficient solution. In Ref. \cite{PhysRevLett.122.193603} a variational formulation is developed and the semianalytical formula is derived for the ground and excited state energies.

\section{Formalism}
We consider two particles with positions $\mathbf{r}_1$, $\mathbf{r}_2$
and charges $q_1$,\ $q_2$. Later we show that an analytical approach only works
for $q_1=q_2$, but it is useful to consider the general case to show the origin of the coupling to the center of mass.
The Hamiltonian of the system is
\begin{equation}
    H=H_e+H_{ph}=H_e+H_p+H_{ep}+H_d.
\end{equation}
$H_e$ is the electronic Hamiltonian,
$H_{ph}$ describes the electron-photon
interaction, which is a sum of three terms, the
photon Hamiltonian $H_p$, the electron-photon coupling $H_{ep}$, and the dipole self-interaction $H_d$.
The electron-photon interaction can be described by using the PF nonrelativistic QED Hamiltonian. The PF Hamiltonian can be rigorously derived
\cite{Ruggenthaler2018,Rokaj_2018,Mandal,acs.jpcb.0c03227,PhysRevB.98.235123} by
applying the Power-Zienau-Woolley gauge transformation \cite{Zienau}, with a unitary phase transformation on the minimal coupling ($p\cdot A$) Hamiltonian in the Coulomb gauge, 
\begin{equation}
    H_{ph}={1\over 2} \sum_{\alpha=1}^{N_p} \left[
p_{\alpha}^2+\omega_\alpha^2\left( q_{\alpha}-{\boldsymbol{\lambda_{\alpha}}\over \omega_\alpha}\cdot\mathbf{D}\right)^2
\right],
\end{equation}
where $\mathbf{D}=\sum_{i=1}^N q_i\mathbf{r}_i$ is the dipole operator. The photon fields are described by quantized oscillators. 
$q_\alpha={1\over \sqrt{2\omega_\alpha}}(\hat{a}^+_\alpha+\hat{a}_\alpha)$ is the displacement field and $p_\alpha=-i\sqrt{{\omega_\alpha}\over 2}(\hat{a}_\alpha-\hat{a}^+_\alpha)$ is the conjugate momentum.
This Hamiltonian describes $N_p$ photon modes with frequency $\omega_{\alpha}$ and coupling  $\boldsymbol{\lambda}_{\alpha}$. The coupling term is usually written as \cite{PhysRevA.90.012508}
\begin{equation}
    \boldsymbol{\lambda}_{\alpha}=\sqrt{4\pi}\,S_\alpha(\mathbf{r})\mathbf{e}_\alpha,
\end{equation}
where $S_\alpha(\mathbf{r})$ is the mode function at position $\mathbf{r}$ and $\mathbf{e}_\alpha$ is the transversal polarization vector of the photon modes.

The three components of the electron-photon interaction are as follows: The photonic part is
\begin{equation}
H_{p}=\sum_{\alpha=1}^{N_p}\left(\frac{1}{2} p_{\alpha}^{2}+\frac{\omega_{\alpha}^{2}}{2} q_{\alpha}^{2}\right) = 
\sum_{\alpha=1}^{N_p} \omega_{\alpha}\left(\hat{a}_{\alpha}^{+} \hat{a}_{\alpha}+\frac{1}{2}\right).
\end{equation}
By using the creation and annihilation operators, the photon states $|n_{\alpha}\rangle$ can be generated by multiple applications of the creation operators on the vacuum state $|n_{\alpha}\rangle=(\hat{a}_{\alpha}^+)^n|0\rangle$. All other photon operations can be done by using $\hat{a}_{\alpha}$ and $\hat{a}_{\alpha}^+$. 
The interaction term is
\begin{equation}
    H_{ep}=-\sum_{\alpha=1}^{N_p}\omega_{\alpha}q_\alpha
    \boldsymbol{\lambda_{\alpha}}\cdot\mathbf{D}=
    -\sum_{\alpha=1}^{N_p}\sqrt{\omega_{\alpha}\over
    2}(\hat{a}_{\alpha}+\hat{a}_{\alpha}^+)\boldsymbol{\lambda_{\alpha}}\cdot\mathbf{D}.
\label{hep}
\end{equation}
Only photon states $|n_{\alpha}\rangle$, $|n_{\alpha}\pm 1\rangle$ are connected by $\hat{a}_{\alpha}$ and $\hat{a}_{\alpha}^+$. The matrix elements 
of the dipole operator $\mathbf{D}$ are only nonzero between spatial basis functions with angular momentum $l$ and $l\pm 1$ in 3D or $m$ and $m\pm 1$ in2D.
The strength of the electron-photon interaction can be characterized by the effective coupling parameter
\begin{equation}
g_{\alpha}=\left|\boldsymbol{\lambda_{\alpha}}\right| \sqrt{\frac{\omega_{\alpha}}{2}}.
\label{str}
\end{equation}
The dipole self-interaction is
\begin{equation}
H_{d}={1\over 2} \sum_{\alpha=1}^{N_p} \left(\boldsymbol{\lambda_{\alpha}} \cdot \mathbf{D}\right)^{2},
\end{equation}
which describes the effects of the polarization of the electrons back on the photon field. The importance of this term for the existence of a ground state is discussed in Ref \cite{Rokaj_2018}.

\subsection{Separation of the relative and center of mass equations}

For simplicity we only consider a single photon mode. The formalism can
be easily extended to many photon modes as described in Appendix \ref{nphoton} and Appendix \ref{lambda}. We will define the coupling 
strength as $\boldsymbol{\lambda}=(\lambda,\lambda,0)$. A more general case is described in Appendix \ref{lambda}. In this section, we consider the Hamiltonian that acts only in the electron space
\begin{eqnarray}
H_e+&H_d&=-\frac{1}{2} \nabla_{1}^{2}+\frac{1}{2} \omega_0^{2} \mathbf{r}_{1}^{2}-\frac{1}{2} \nabla_{2}^{2}+\frac{1}{2} \omega_0^{2} \mathbf{r}_{2}^{2}\nonumber \\ &+&\frac{q_1 q_2}{\left|\mathbf{r}_{1}-\mathbf{r}_{2}\right|} +{1\over 2} (q_1\boldsymbol{\lambda}\cdot\mathbf{r}_1+
q_2\boldsymbol{\lambda}\cdot\mathbf{r}_2)^2.\ 
\end{eqnarray}
 Atomic units $ \hbar=m=e=1 $ are used throughout and unit charges are assumed. 

Defining relative and CM coordinates as
\begin{equation}
\begin{array}{l}
\mathbf{r}=\mathbf{r}_{2}-\mathbf{r}_{1}, \\
\mathbf{R}=\frac{1}{2}\left(\mathbf{r}_{1}+\mathbf{r}_{2}\right),
\end{array}
\end{equation}
the Hamiltonian decouples into a relative and CM Hamiltonian
\begin{eqnarray}
H_e+H_d&=&-\nabla_{\mathbf{r}}^{2}+\frac{1}{4} \omega_0^{2} \mathbf{r}^{2}+\frac{q_1q_2}{r}-\frac{1}{4} \nabla_{\mathbf{R}}^{2}+\omega_0^{2} \mathbf{R}^{2}\nonumber \\
&+&
{1\over 2}\left(\boldsymbol{\lambda}\cdot\left((q_1+q_2)\mathbf{R}+{1\over 2} (q_1-q_2)\mathbf{r}\right)\right)^2
\nonumber\\
&\equiv& H_{\mathbf{r}}+H_{\mathbf{R}},\label{hant}
\end{eqnarray}
and the corresponding eigenvalue problem is
\begin{equation}
    \left(H_{\mathbf{r}}+H_{\mathbf{R}}\right)\Phi(\mathbf{r},\mathbf{R})=(\epsilon+\eta)\Phi(\mathbf{r},\mathbf{R}),
\end{equation}
and $E=\epsilon+\eta$ is the eigenenergy. Note that for like charges, the last term only contributes to $H_{\mathbf{R}}$,
otherwise, it only contributes to $H_{\mathbf{r}}$ and  there is no cross term between $\mathbf{R}$ and $\mathbf{r}$.
\subsubsection{$q_1=-q_2$}
In this case the photon only couples to $\mathbf{r}=(x,y,z)$. The CM
wave function is a harmonic oscillator eigenfunction with frequency $2\omega_0$. By introducing $u={x+y\over\sqrt{2}}$, and  $v={-x+y\over\sqrt{2}}$, the relative motion Hamiltonian takes the form
\begin{equation}
\begin{aligned}
    H_{\mathbf{r}}=-&\nabla_u^2-\nabla_v^2-\nabla_z^2+{1\over 2} \omega_u^2 u^2+{1\over 2} \omega_v^2 v^2+{1\over 2} \omega_z^2 z^2\\&-{1\over {(u^2+v^2+z^2)}^{1/2}},
\end{aligned}
\end{equation}
where ${\omega_u}^2=2\lambda^2+\frac{1}{2}{\omega_0}^2,\ {\omega_v}^2={\omega_z}^2=\frac{1}{2}{\omega_0}^2$. This is a single particle Coulomb problem in an anisotropic harmonic potential. The derivation is detailed in the next section. We are
not aware of any existing analytical solutions to this system. One can, in principle, solve this problem using a product basis of the $u-v-z$ harmonic oscillators, but we do not pursue this case any further in this paper.

\subsubsection{$q_1=q_2$}
\label{like}
In the following, we will consider $q_1=q_2$ because, in this case, the
equation for the relative motion can be analytically found for
certain frequencies as mentioned before. After multiplying the relative part by 1/2 and the CM part by 2 to bring the equations 
in a more convenient form, we have
\begin{equation}\label{equ1}
\left[-\frac{1}{2} \nabla_{\mathbf{r}}^{2}+\frac{1}{2} \omega_{\mathbf{r}}^{2} \mathbf{r}^{2}+\frac{1}{2} \frac{1}{r}\right] \varphi(\mathbf{r})=\varepsilon^{\prime} \varphi(\mathbf{r}),
\end{equation}
where $ \omega_{\mathrm{r}}=\frac{1}{2} \omega_0 $ and $ \varepsilon^{\prime}=\frac{1}{2} \varepsilon $, and
\begin{equation}\label{equ2}
\left[-\frac{1}{2} \nabla_{\mathbf{R}}^{2}+\frac{1}{2} \omega_{\mathbf{R}}^{2} \mathbf{R}^{2}+4(\boldsymbol{\lambda}\cdot\mathbf{R})^2
\right] \xi(\mathbf{R})=\eta^{\prime} \xi(\mathbf{R}),
\end{equation}
where $ \omega_{\mathbf{R}}=2 \omega_0 $ and $ \eta^{\prime}=2 \eta $.
The total wave function can be written as
\begin{equation}
\Phi(\mathbf{r},\mathbf{R})=
\varphi(\mathbf{r})\xi(\mathbf{R}).
\end{equation}
In this case, the CM motion in the $z$-direction is described by a harmonic oscillator eigenfunction, and we drop this part from now. 

In 2D, using $\mathbf{R}=(X,Y)$ one can rewrite $H_{\mathbf{R}}$
as (in 3D one simply has to multiply the CM wave 
function with a harmonic oscillator function of frequency $2\omega_0$ in the $Z$
direction)
\begin{equation}\label{equ3}
H_{\mathbf{R}}=
-{1\over 2}{\partial^2 \over \partial X^2}
-{1\over 2}{\partial^2 \over \partial Y^2}
+{1\over 2}\omega_X^2 X^2+{1\over 2}\omega_Y^2 Y^2 +{1\over 2}\omega_{XY}^2 XY,
\end{equation}
where 
\begin{equation}\label{equ4}
    \omega_X^2=\omega_Y^2=\omega_{\mathbf{R}}^2+8\lambda^2, \ \ \ \ 
\omega_{XY}^2=16\lambda^2.
\end{equation}
Using a unitary transformation (a generalized version is presented
in Appendix \ref{nphoton})
\begin{equation}
    U={X+Y\over\sqrt{2}}, \ \ \ \ V={-X+Y\over\sqrt{2}},
\end{equation}
we have
\begin{eqnarray}\label{harm}
H_{\mathbf{R}}&=&
-{1\over 2}{\partial^2 \over \partial U^2}
-{1\over 2}{\partial^2 \over \partial V^2}
+{1\over 2}\omega_U^2 U^2+{1\over 2}\omega_V^2 V^2\nonumber \\
&\equiv&  H_U+H_V,
\end{eqnarray}
where 

\begin{equation}
    \omega_U^2=\frac{1}{2}(\omega_X^2+\omega_{XY}^2
    +\omega_Y^2)=\omega_R^2+16\lambda^2,
\end{equation} 
\begin{equation}
    \omega_V^2=\frac{1}{2}
    (\omega_X^2-\omega_{XY}^2+\omega_Y^2)=\omega_R^2.
\end{equation}

This Hamiltonian is  analytically solvable: the lowest energy is
\begin{equation}
    \eta={1\over 2} \left(\omega_0+\sqrt{\omega_0^2+4\lambda^2}\right).
\end{equation}
$H_{\mathbf{r}}$ is also analytically solvable, in this case, for
certain frequencies \cite{taut,Supp}. For example,  for $\omega_0=1$
one gets $\epsilon=2$ (see the Table in Ref. \cite{Supp}) and the total energy
is $E=2+{1\over 2}+{1\over 2} \sqrt{1+4\lambda^2}$. 

The wave function of the CM motion now can be written as
\begin{equation}
\xi(\mathbf{R})=\phi_k({U})\phi_l(V),
\end{equation}
where $\phi_k$ is the $k$th eigenfunction of the one-dimensional
harmonic oscillator,
\begin{equation}
\phi_k(U)=\left(\sqrt{\omega_U}\over \sqrt{\pi}\,2^k k!\right)^{1\over 2} {\rm e}^{-{\omega_U\over 2}U^2}
H_k(\sqrt{\omega_U}\,U),
\end{equation}
and the eigenfunctions are similarly defined for $V$.

\subsection{Photon-electron coupling}
The coupling term Eq. \eqref{hep} takes the form
\begin{equation}
H_{ep}=-\sqrt{{\omega\over 2}}(\hat{a}+\hat{a}^+)  \lambda D, \ \ \ \  D=2\sqrt{2} U,
\end{equation}
so only the $U$ harmonic oscillators are coupled with photons. The Hamiltonian that we have to solve is reduced to a single one-dimensional electronic Hamiltonian coupled to light:
\begin{equation}
H_c=H_U+\omega \left(\hat{a}^+\hat{a}+{1\over 2}\right)-2\omega \sqrt{2} \lambda U q.
\end{equation}

This Hamiltonian can be solved by exact diagonalization using the basis states 
\begin{equation}
    \phi_k(U)|n\rangle.
\end{equation}
For the diagonalization, one needs the matrix elements of the Hamiltonian
which are readily available. The operators $H_U$ and $U$ act on the real space, and $\hat{a}+\hat{a}^+$ acts on the photon space.
For the coupling term in the photon space:
\begin{eqnarray}
q|n\rangle&=&\frac{1}{\sqrt{2
\omega}}\left(\hat{a}+\hat{a}^{+}\right)|n\rangle\\
&=&
\frac{1}{\sqrt{2 \omega}}\left(|\sqrt{n}| n-1\rangle+\sqrt{n+1}|n+1\rangle\right) \nonumber,
\end{eqnarray}
and the matrix elements of $q$ are
\begin{equation}
\langle m\vert q\vert n\rangle= \frac{1}{\sqrt{2\omega}} D_{mn},
\end{equation}
where
\begin{equation}
D_{mn}=\left(\begin{array}{cccccc}0 & \sqrt{1} & 0 & 0 & 0 & \ldots \\ \sqrt{1} & 0 & \sqrt{2} & 0 & 0 & \ldots \\ 0 & \sqrt{2} & 0 & \sqrt{3} & 0 & \ldots \\ 0 & 0 & \sqrt{3} & 0 & \sqrt{4} & \ldots \\ 0 & 0 & 0 & \sqrt{4} & 0 & \ldots \\ \vdots & \vdots & \vdots & \vdots & \vdots & \ddots\end{array}\right) .
\end{equation}
The Hamiltonian $H_U$ is diagonal in the harmonic oscillator bases
\begin{equation}
    \langle \phi_i\vert H_U \vert\phi_j\rangle=(j+{1\over 2} )\omega_U\delta_{ij}.
\end{equation}
The matrix elements of the photon Hamiltonian are
\begin{equation}
    \langle n\vert \omega\left(\hat{a}^+\hat{a}+{1\over 2}\right) \vert m\rangle=(n+{1\over 2} )\omega \delta_{nm}.
\end{equation}

The last piece is the matrix elements of the position operator in harmonic oscillator bases:
\begin{equation}
    \langle \phi_i\vert U\vert\phi_j\rangle=\frac{1}{\sqrt{2\omega_U}}D_{ij}.    
\end{equation}

Thus, the matrix elements of $H_c$ are
\begin{widetext}
\begin{equation}\label{matr}
    \langle m,\phi_i\vert H_c\vert n,\phi_j\rangle=
    \delta_{mn}\delta_{ij}(j+\frac{1}{2})\omega_U+ \delta_{mn}\delta_{ij}(n+\frac{1}{2})\omega+
    \sqrt{\frac{2\omega}{\omega_U}}\,\lambda\, D_{mn} D_{ij}.
\end{equation}
\end{widetext}
This is a very sparse matrix and can be diagonalized with sparse matrix approaches even for very large dimensions. In practice, a few dozen photon bases $|n\rangle$ and harmonic oscillator bases $\phi_i$ give 
converged energies. This matrix is generalized for $N_p$ photon modes in Appendix \ref{nphoton}. 

After the diagonalization, we have the eigenenergies $\eta_j'$ and the 
eigenfunctions by defining the spatial wave function in the photon 
subspace $n$ as
\begin{equation}
    \psi_j(\mathbf{R})=\phi_0(V)\phi_j(U),
\end{equation}
the eigenfunction for the CM motion is
\begin{eqnarray}
\xi_k(\mathbf{R})&=&\sum_{n=0}^{K_n}\left(\sum_{j=0}^{K_U} c^k_{j,n}\psi_n(\mathbf{R})\right)|n\rangle\nonumber\\&=&
\sum_{j=0}^{K_U}\left(\sum_{n=0}^{K_n} c^k_{j,n}|n\rangle\right)\psi_j(\mathbf{R}),
\label{trunc}
\end{eqnarray}
where $K_U$ and $K_n$ are some suitably chosen upper limits that control the convergence of the eigenvalues. For the $V$ part of the CM motion, we have chosen the lowest state. The first line
in Eq. \eqref{trunc} emphasizes the coupling of the spatial part to
photon spaces; the second line emphasizes the coupling of the linear combination of photon states to a given CM eigenfunction.

\section{Results and discussion}
A few examples will be presented in this section. For these calculations, we have picked an oscillator frequency $\omega_0$ from the Table of Ref. \cite{Supp}, calculated the radial part of the relative wave function as described in Refs. \cite{Supp,taut,taut2}, and multiplied with the corresponding spherical function. This function is then multiplied by $\xi_k(\mathbf{R})$ calculated using Eq. \eqref{trunc}. 

First, we show the wave function for the different CM excitation. In this case, two variables determine the behavior: the confining strength $\omega_0$ and the coupling parameter $\lambda$. We show 2D examples because they are 
easier to visualize. The 3D cases are very similar, with the only difference being that the wave function is multiplied by the lowest harmonic oscillator function with frequency $2\omega_0$ in the $Z$
direction.

\begin{figure}
\begin{minipage}{0.5\textwidth}
\includegraphics[width=0.45\textwidth]{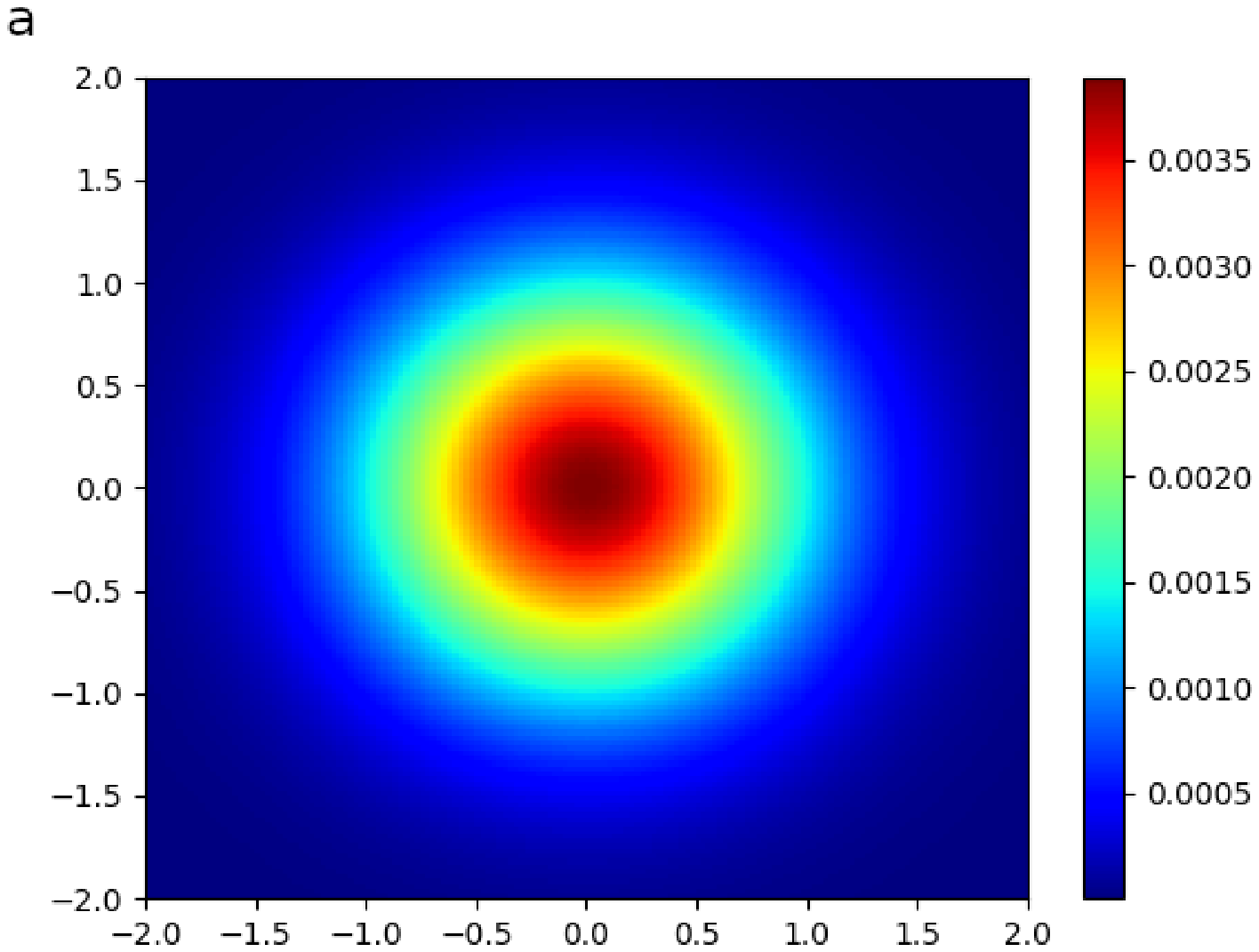}\quad
\includegraphics[width=0.45\textwidth]{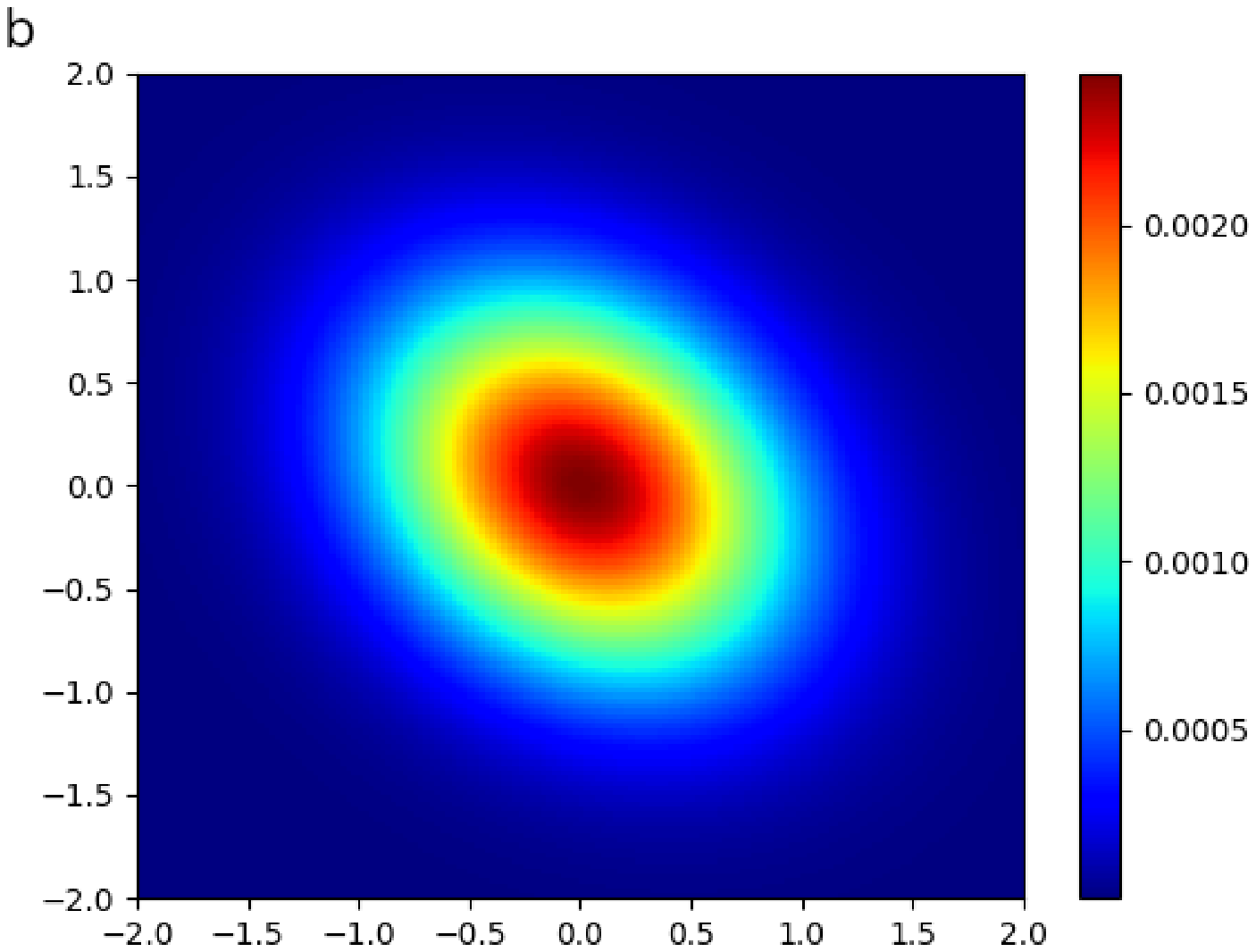}\\\vskip 1cm
\includegraphics[width=0.45\textwidth]{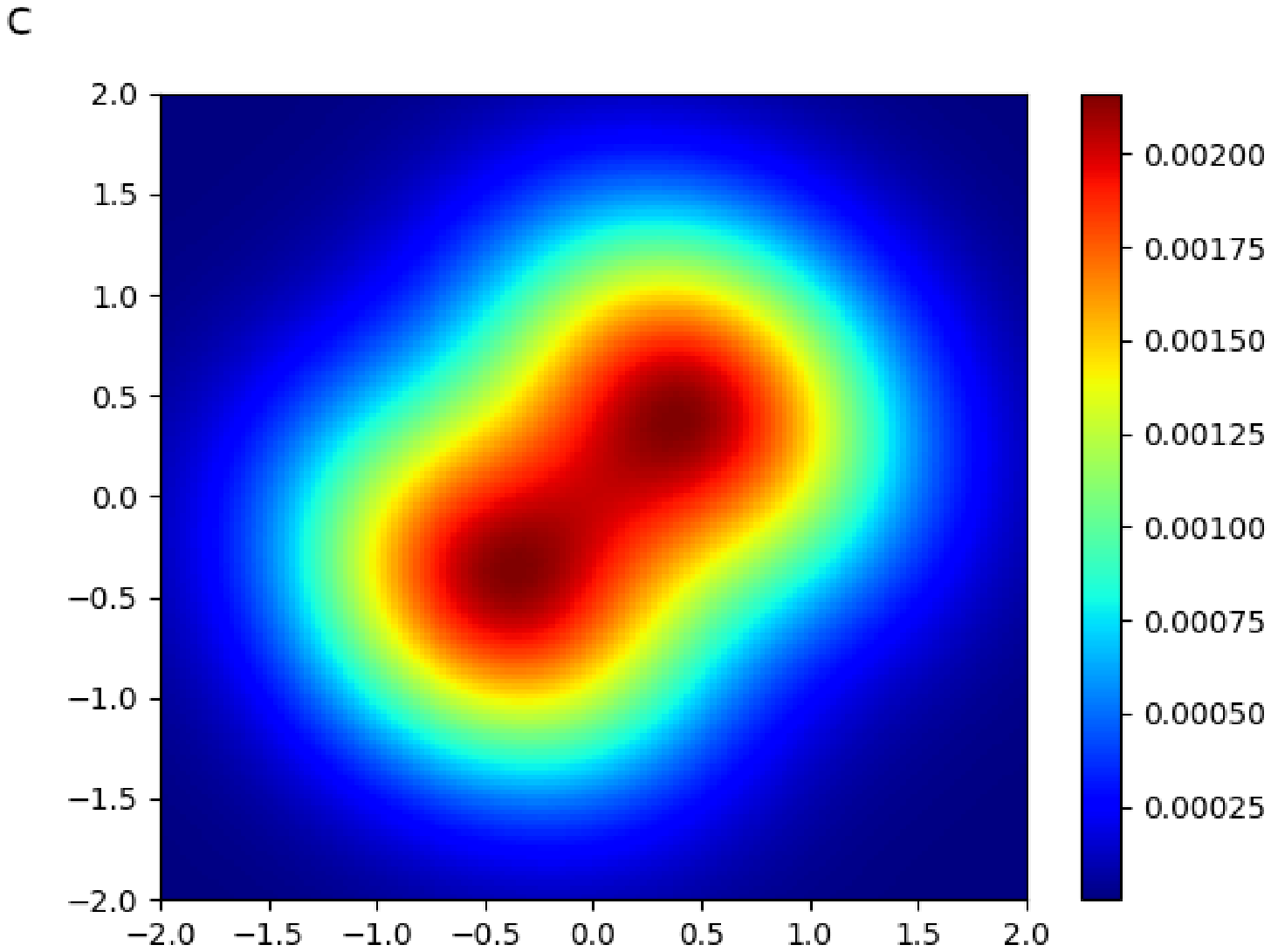}\quad
\includegraphics[width=0.45\textwidth]{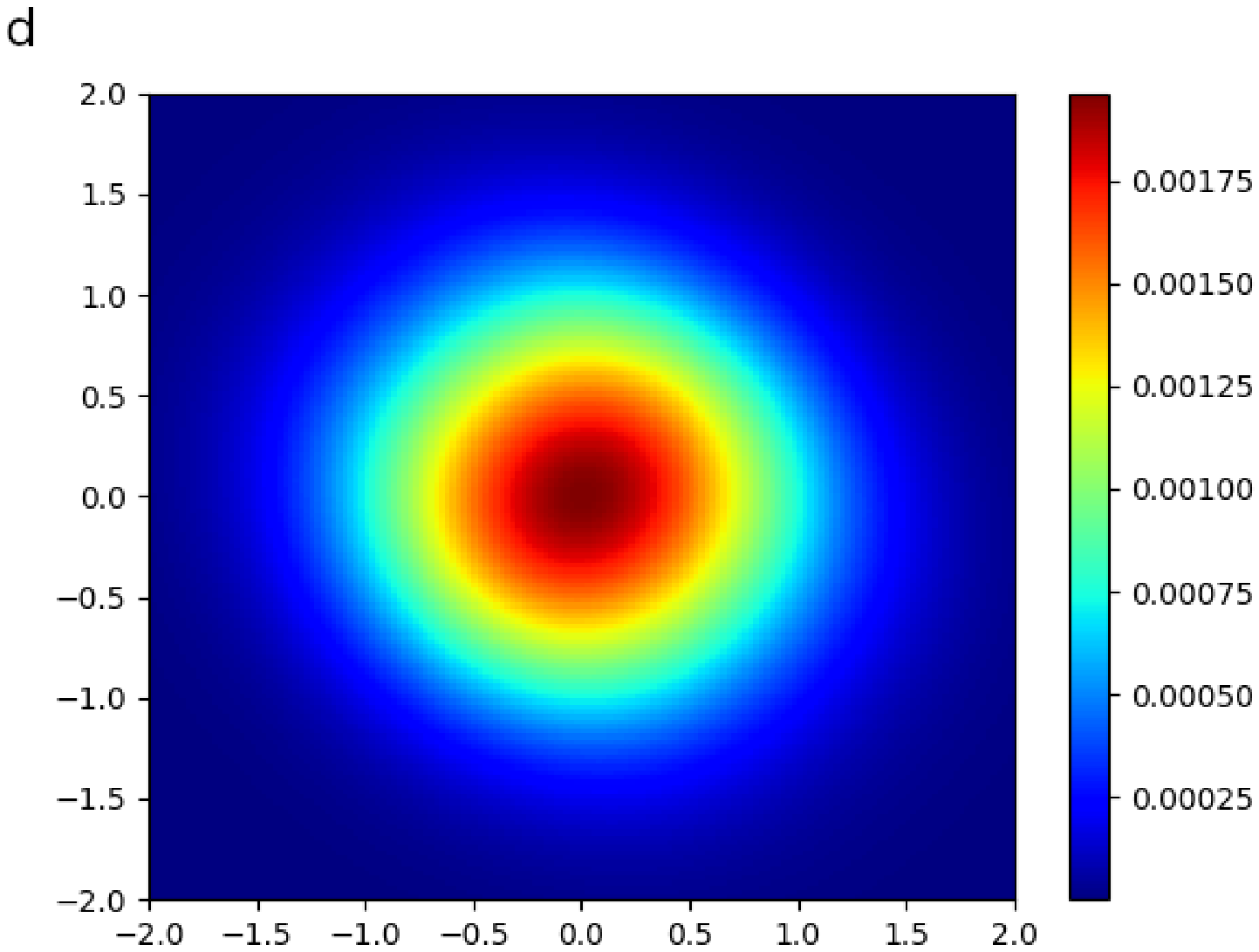}\\
\includegraphics[width=0.45\textwidth]{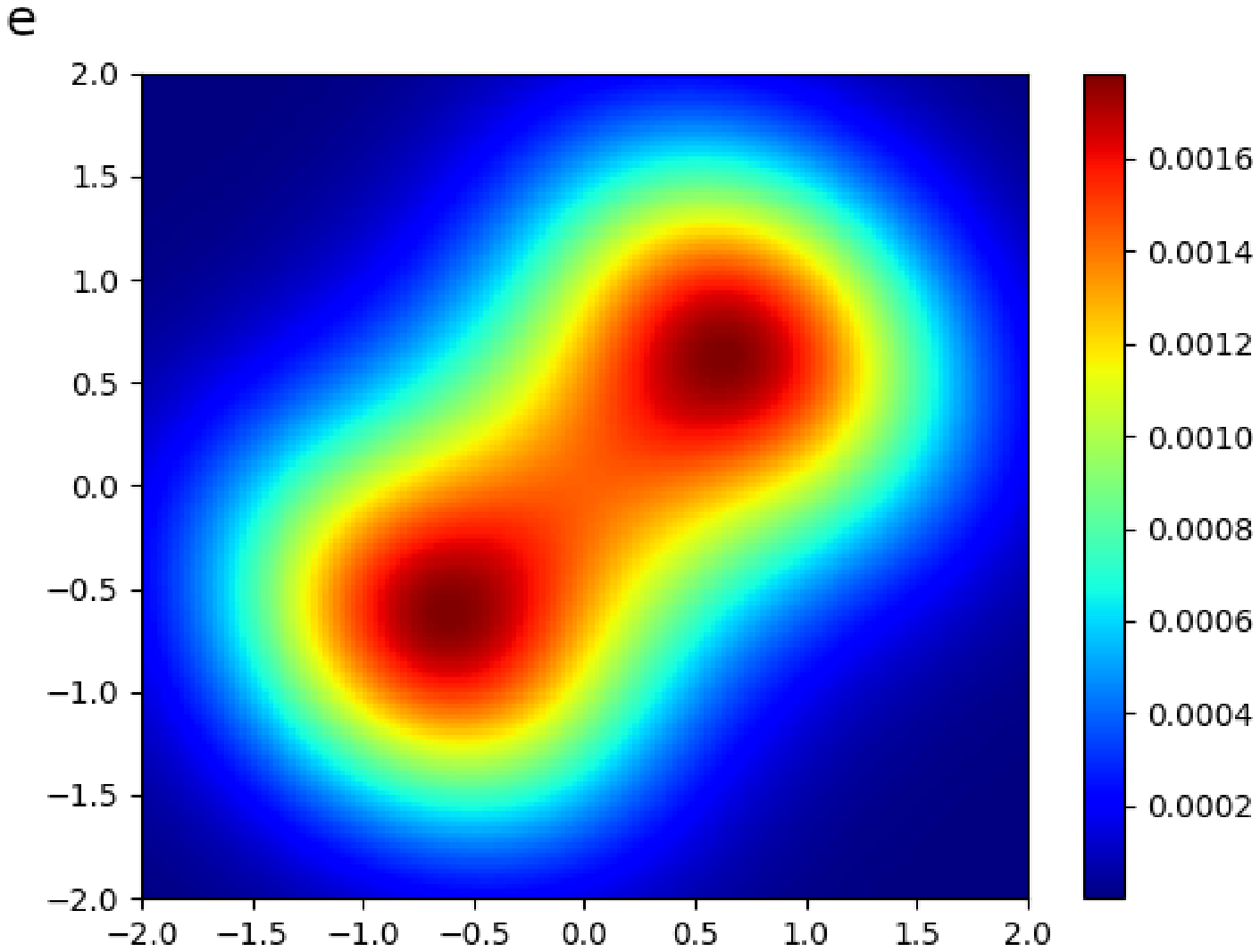}\quad
\includegraphics[width=0.45\textwidth]{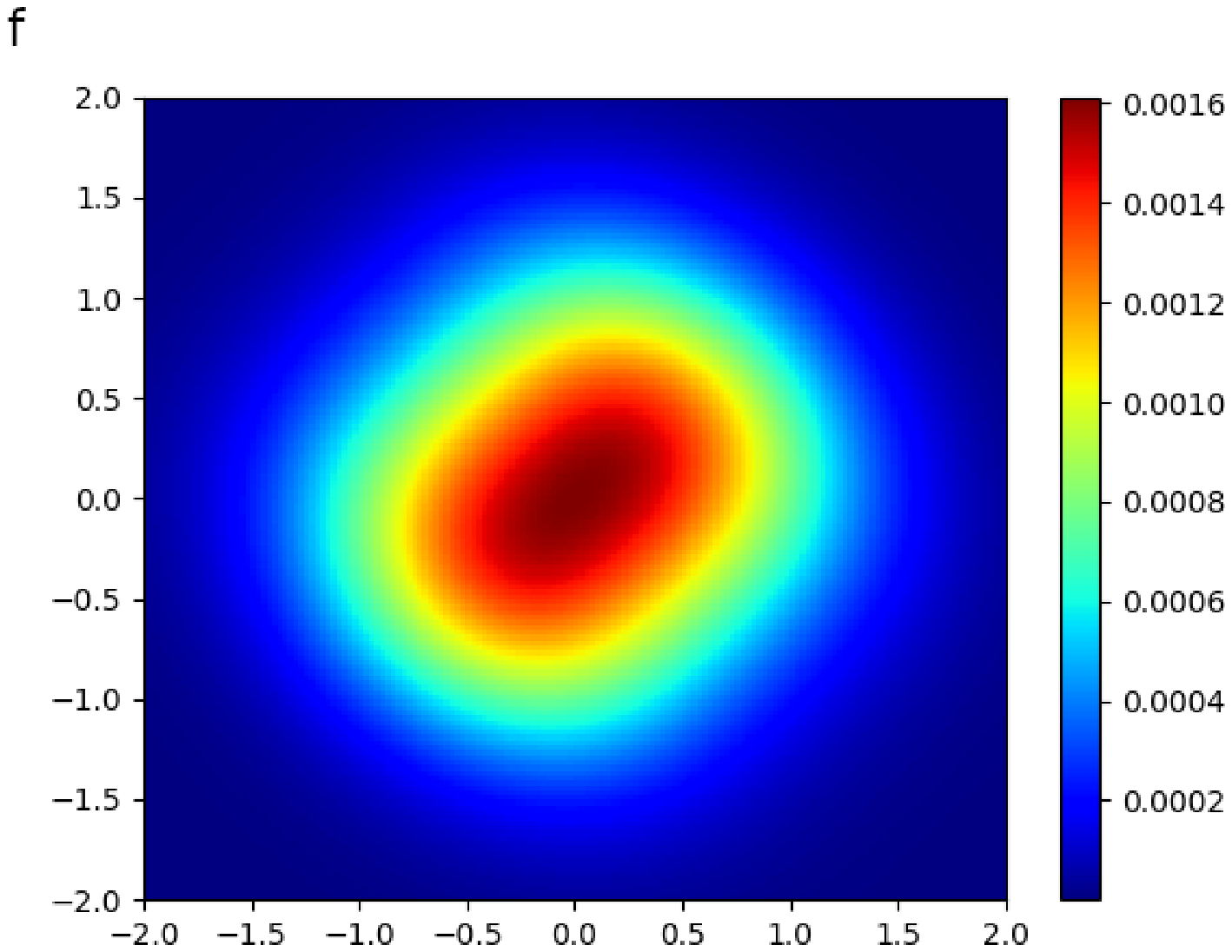}\\\vskip 1cm
\includegraphics[width=0.45\textwidth]{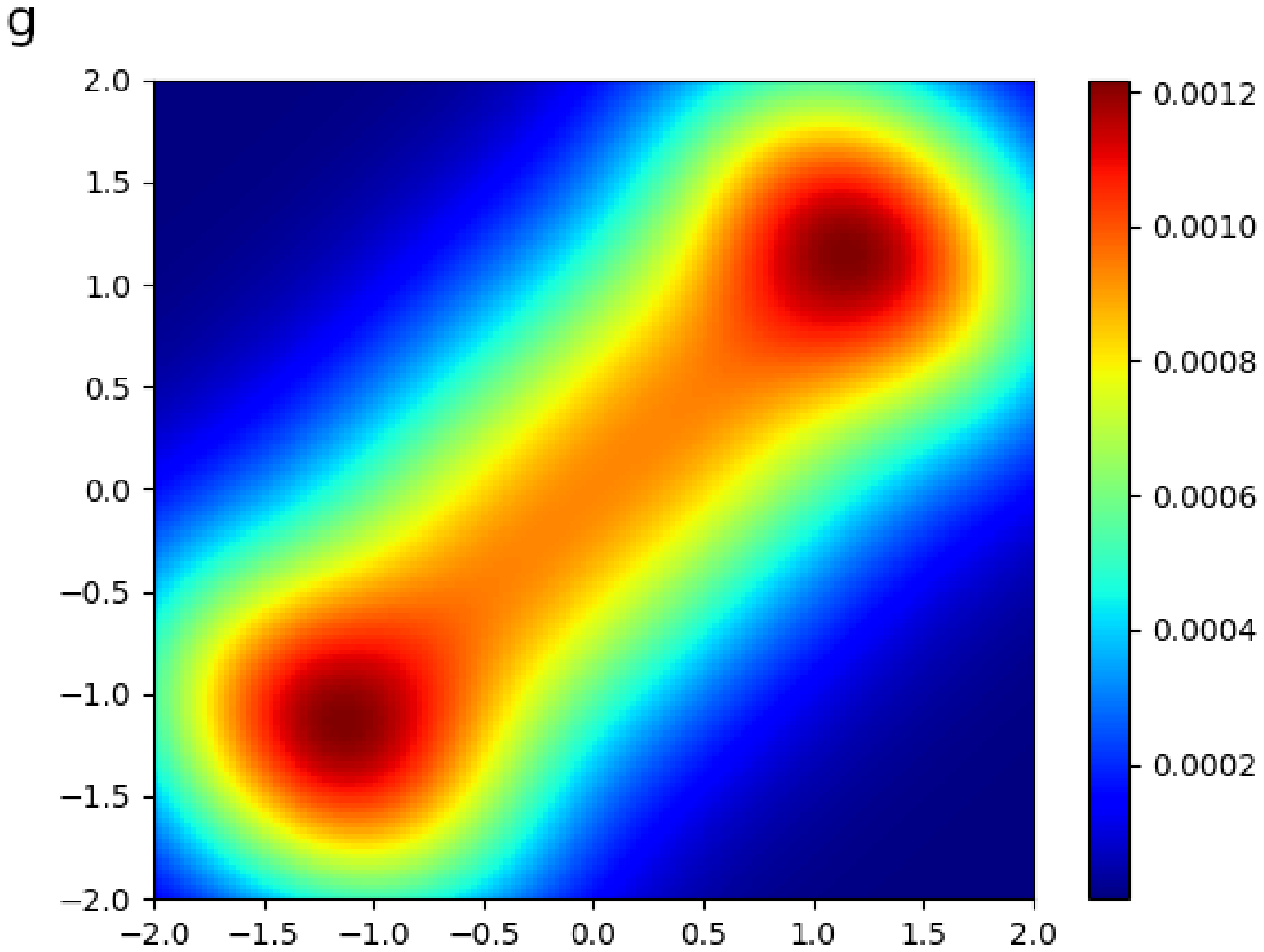}\quad
\includegraphics[width=0.45\textwidth]{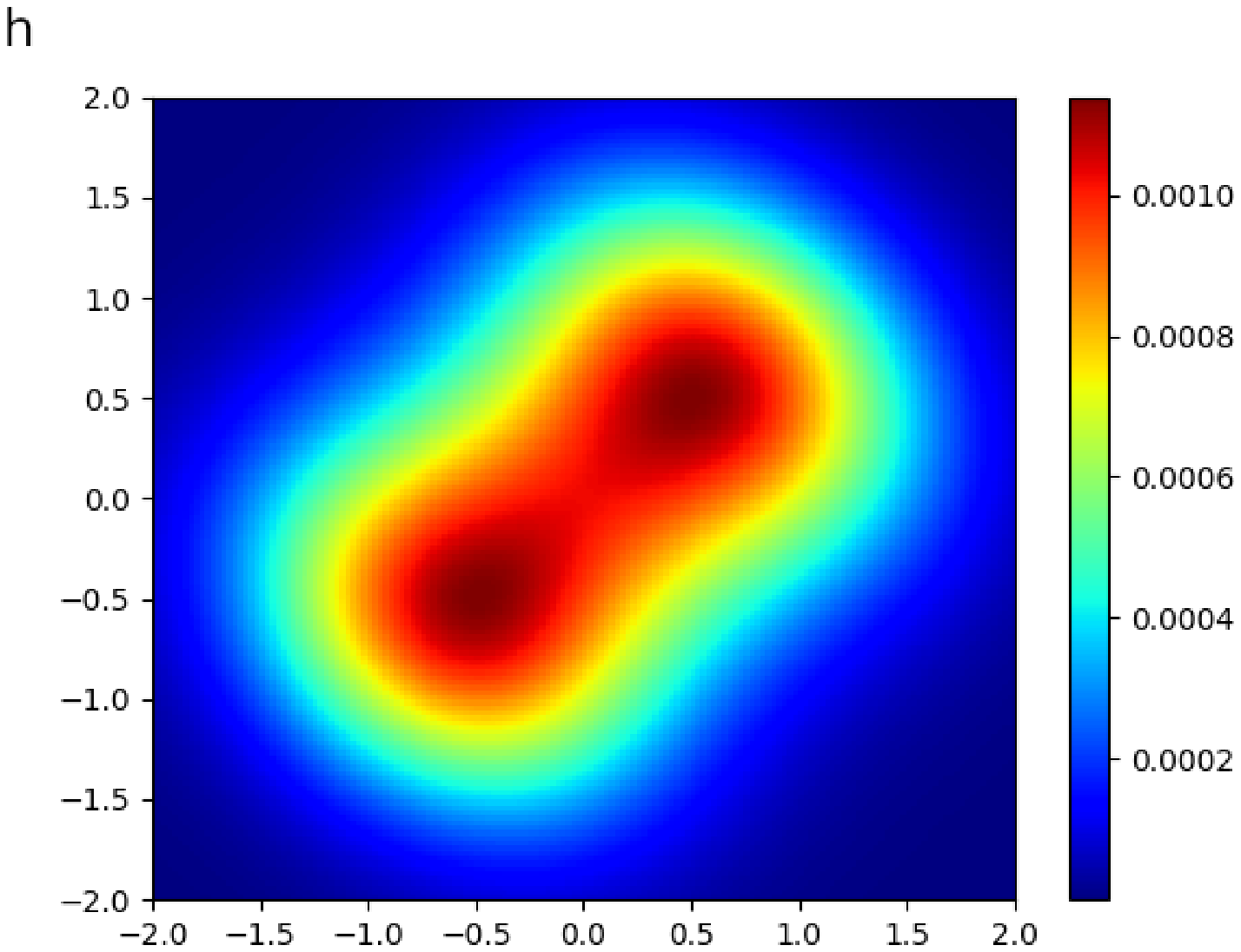}
\end{minipage}\\[1em]
\caption{Two-dimensional densities $\left(\phi(\mathbf{r})\psi_j(\mathbf{R})\right)^2$  of two electrons confined by a harmonic potential with
$\omega_0=1$ a.u. and spin $S=0$. 
First row:  j=0, (a) $\lambda=0.5$, (b) $\lambda=2$.
Second row: j=1, (c) $\lambda=0.5$, (d) $\lambda=2$.
Third  row: j=2, (e) $\lambda=0.5$, (f) $\lambda=2$.
Fourth row: j=5, (g) $\lambda=0.5$, (h) $\lambda=2$.}
\label{singlet}
\end{figure}

First, we show the spin-singlet case using 
$\omega_0=1$. The energy of the relative motion is $\epsilon=1$ a.u.
in this case (see the Table in Ref. \cite{Supp}). The state with
$j=0$ CM wave function is spherically symmetric for
small $\lambda$ (Fig. \ref{singlet}a), and as the anharmonicity 
of the harmonic oscillator dominates ($\omega_V<<\omega_U$), a slightly ellipsoidal
structure appears (Fig. \ref{singlet}b). For $j=1$, the CM
state is multiplied by $U$ ($H_1(\sqrt{\omega_U}\,U)=2\sqrt{\omega_U}\,U$)
and becomes elongated in the diagonal direction (Fig. \ref{singlet}c). This direction is set by the choice of $\boldsymbol{\lambda}=(\lambda,\lambda$), and other
values would change the direction (see Appendix \ref{nphoton}). For larger 
$\lambda$, the 
confinement by $\omega_U$ is much stronger and the elongation disappears (Fig. \ref{singlet}d). For higher $j$ values the 
elongation increases due to the higher $H_j(\sqrt{\omega_U}U)$
polynomials (Figs. \ref{singlet}e and \ref{singlet}g). Higher $\lambda$ values
decrease the elongation (Figs. \ref{singlet}f and \ref{singlet}h). This trend
continues for even higher $j$ values as well. Solutions with other $\omega_0$ values show very similar behaviors. 

\begin{center}
\begin{figure*}
\begin{minipage}{0.95\textwidth}
\includegraphics[width=0.3\textwidth]{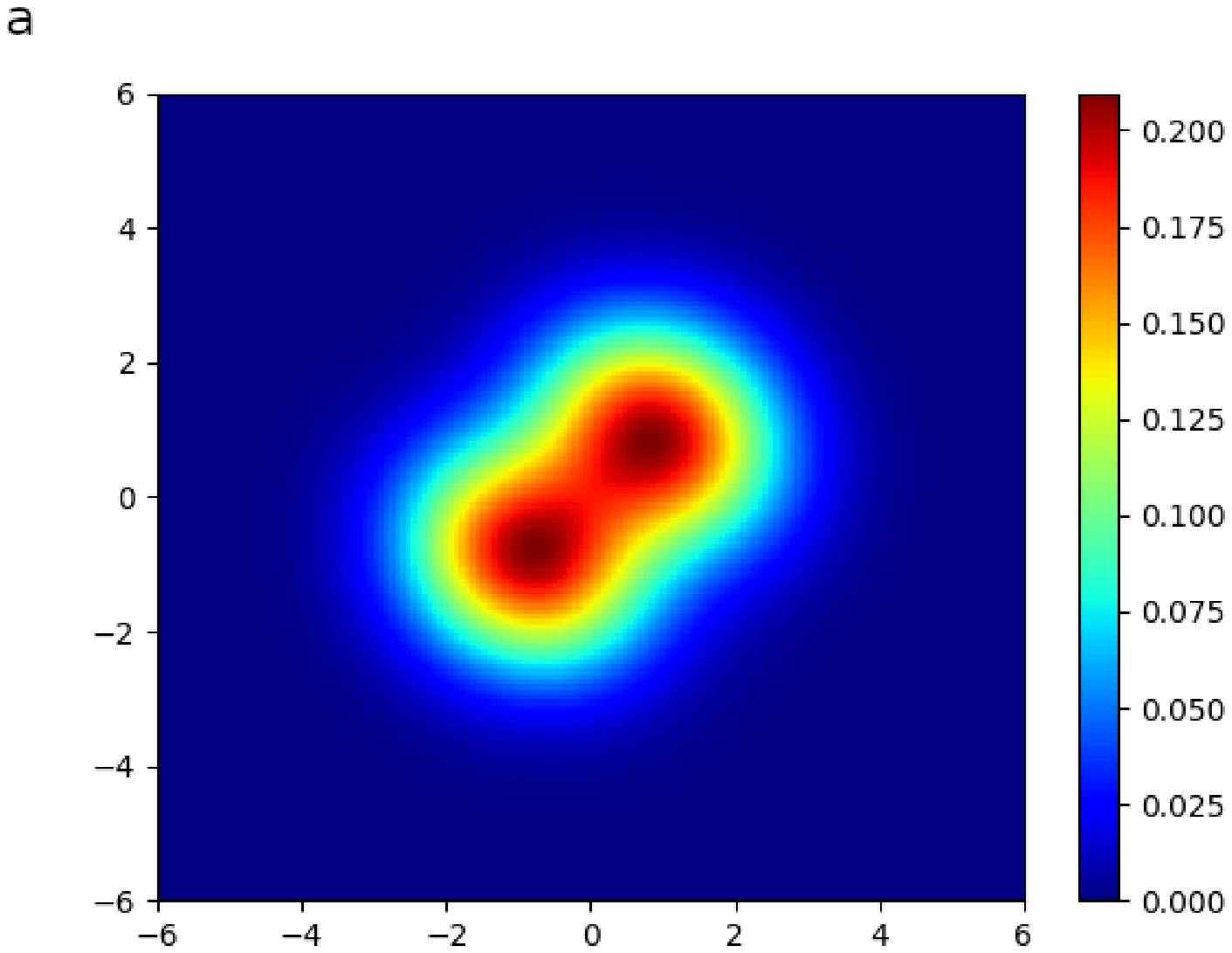}\quad
\includegraphics[width=0.3\textwidth]{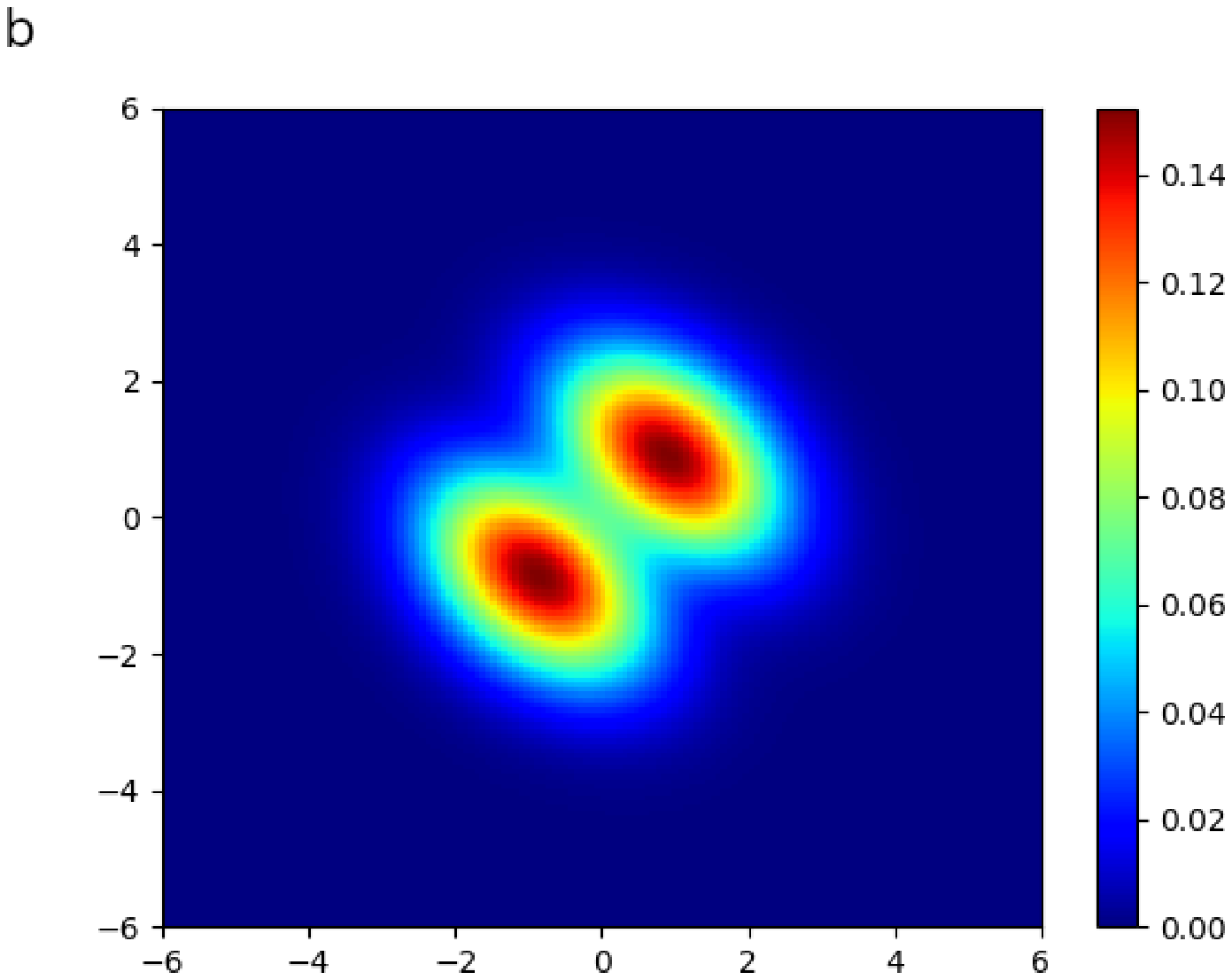}\quad
\includegraphics[width=0.3\textwidth]{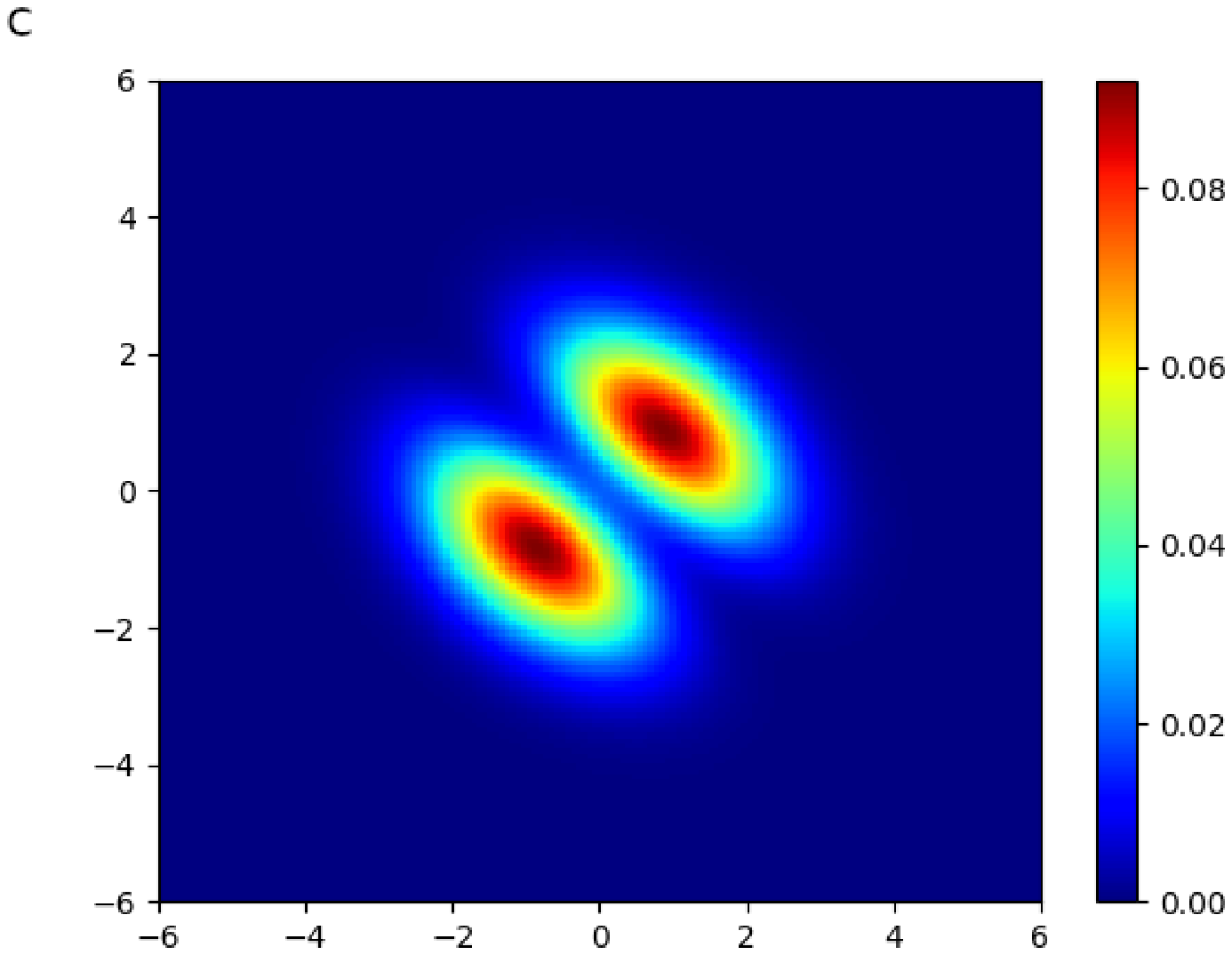}\\\vskip 1cm
\includegraphics[width=0.3\textwidth]{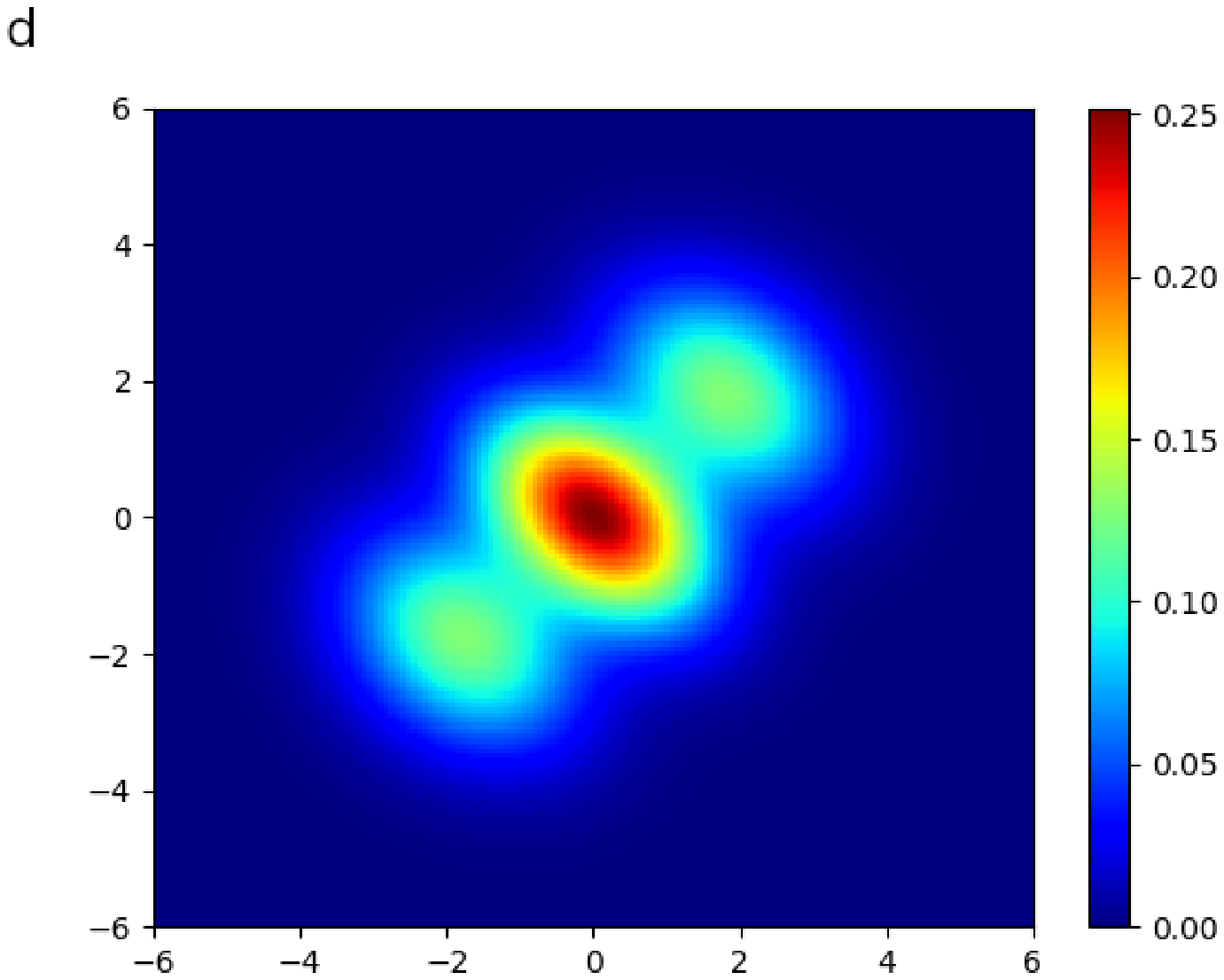}\quad
\includegraphics[width=0.3\textwidth]{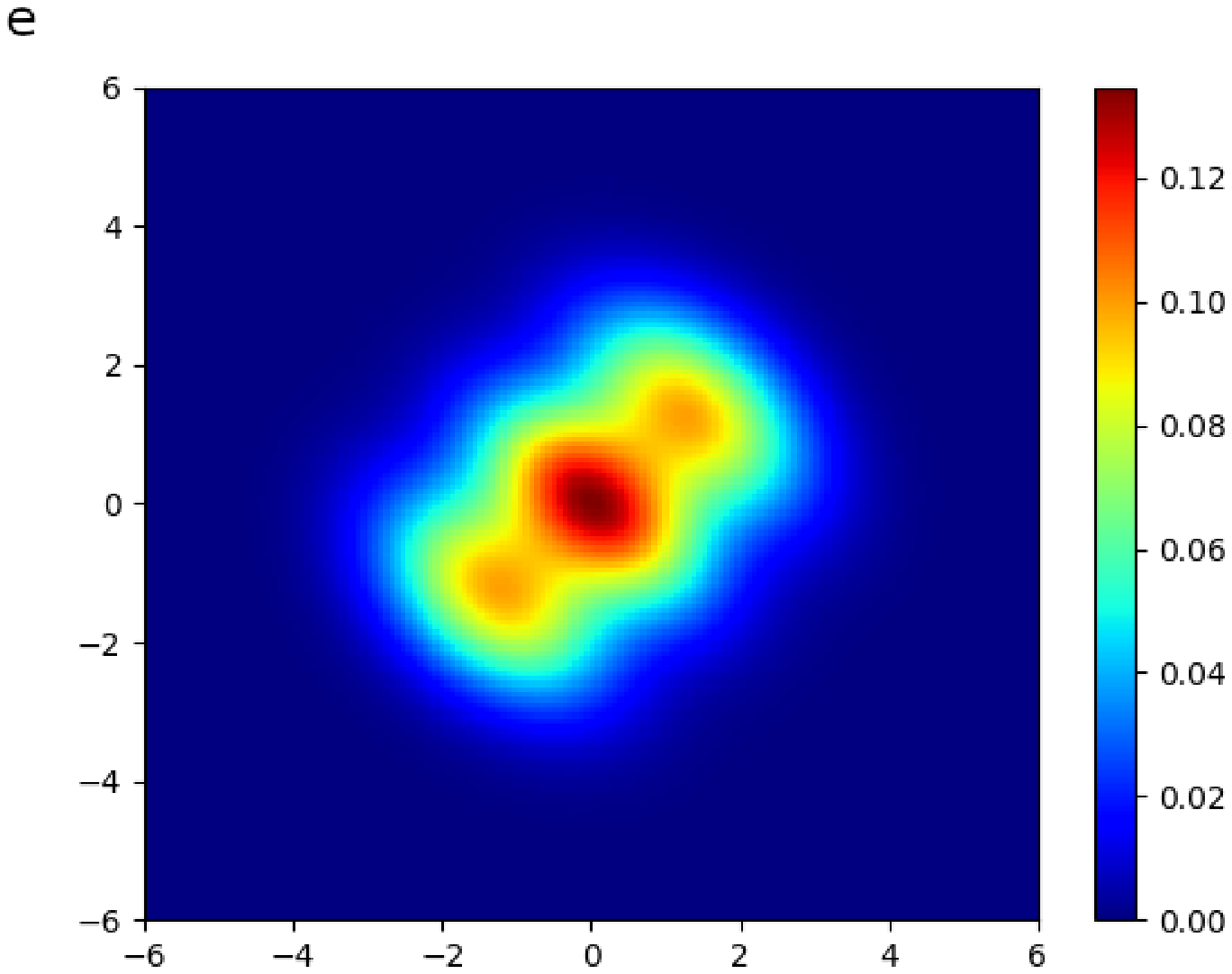}\quad
\includegraphics[width=0.3\textwidth]{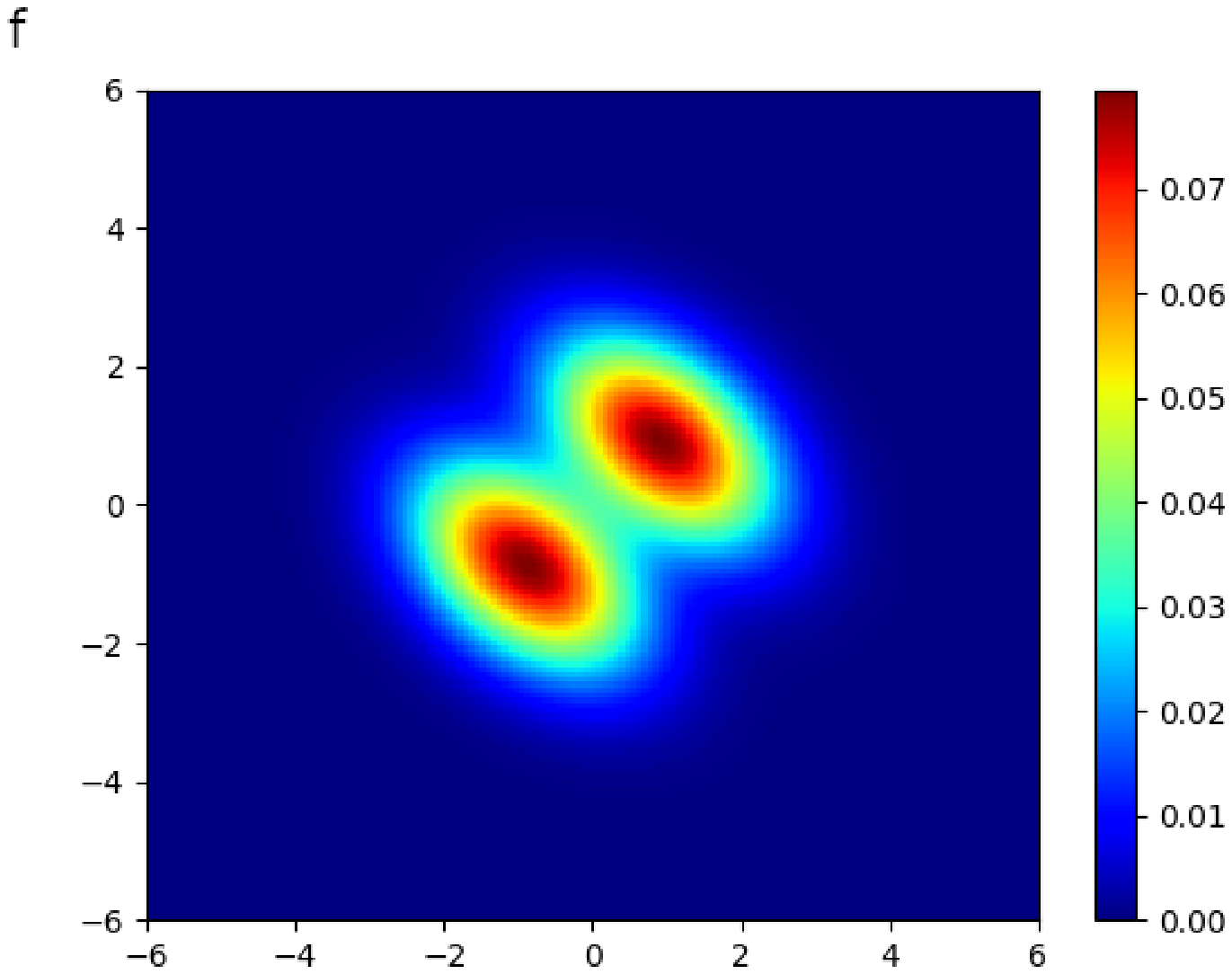}\\\vskip 1cm
\includegraphics[width=0.3\textwidth]{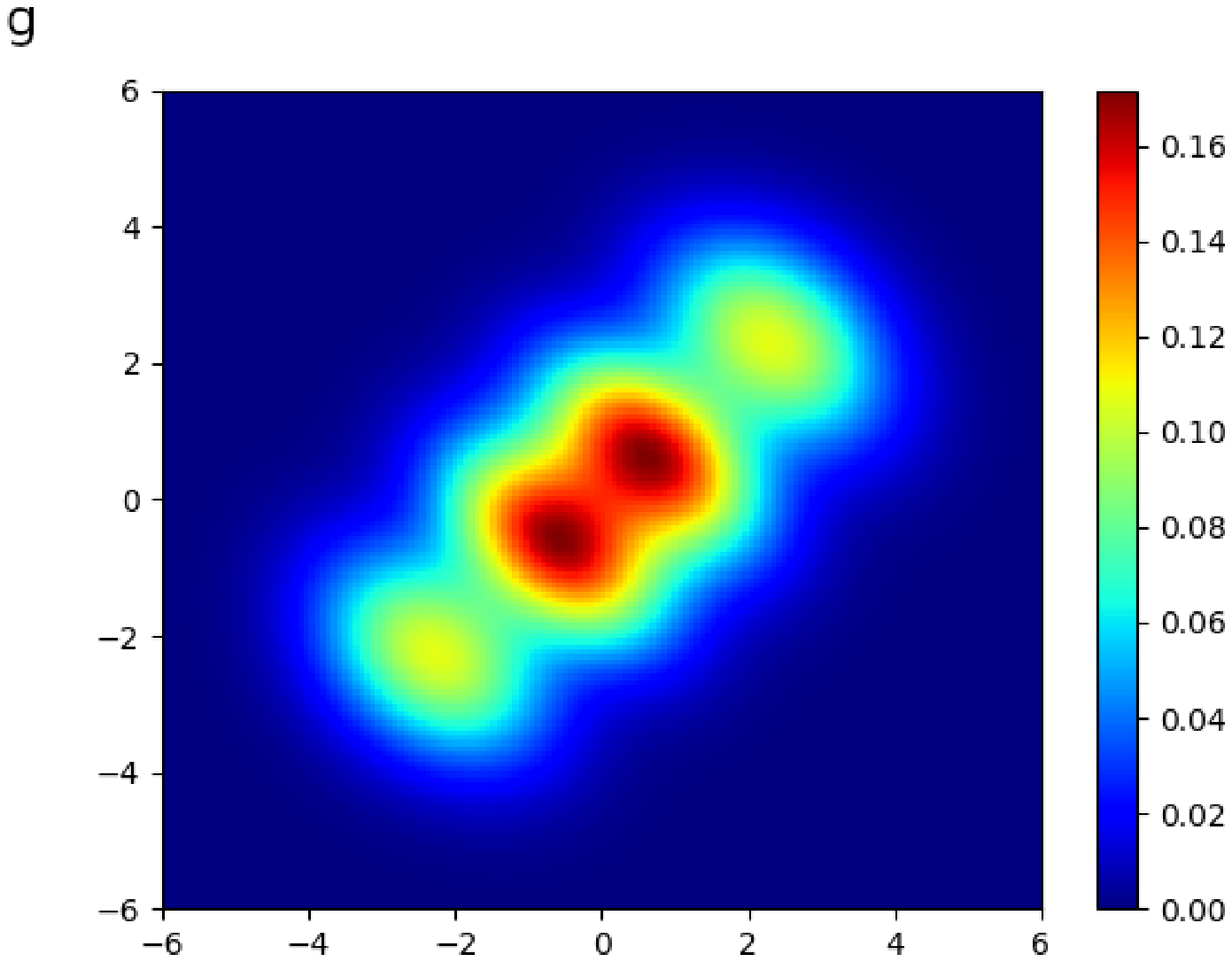}\quad
\includegraphics[width=0.3\textwidth]{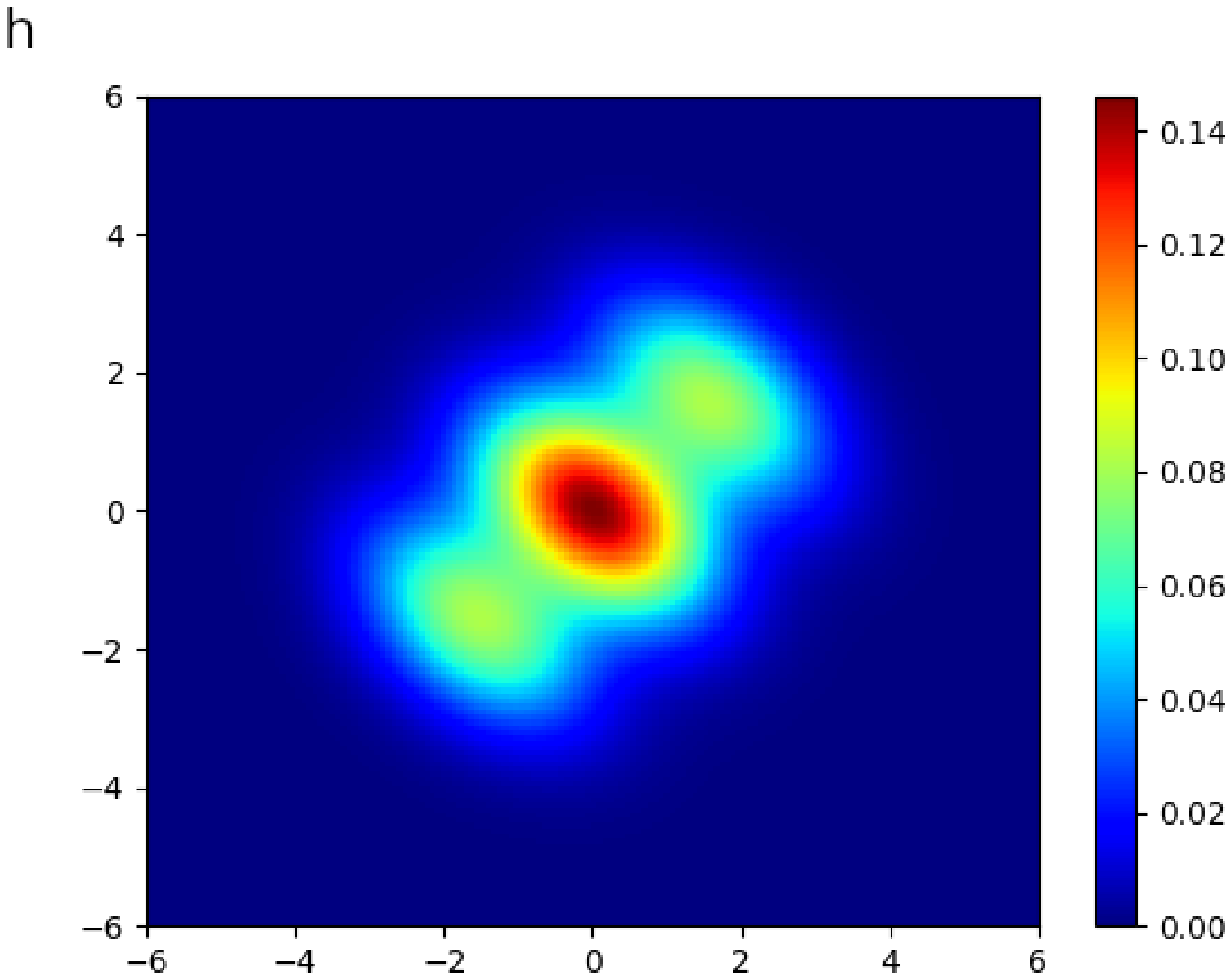}\quad
\includegraphics[width=0.3\textwidth]{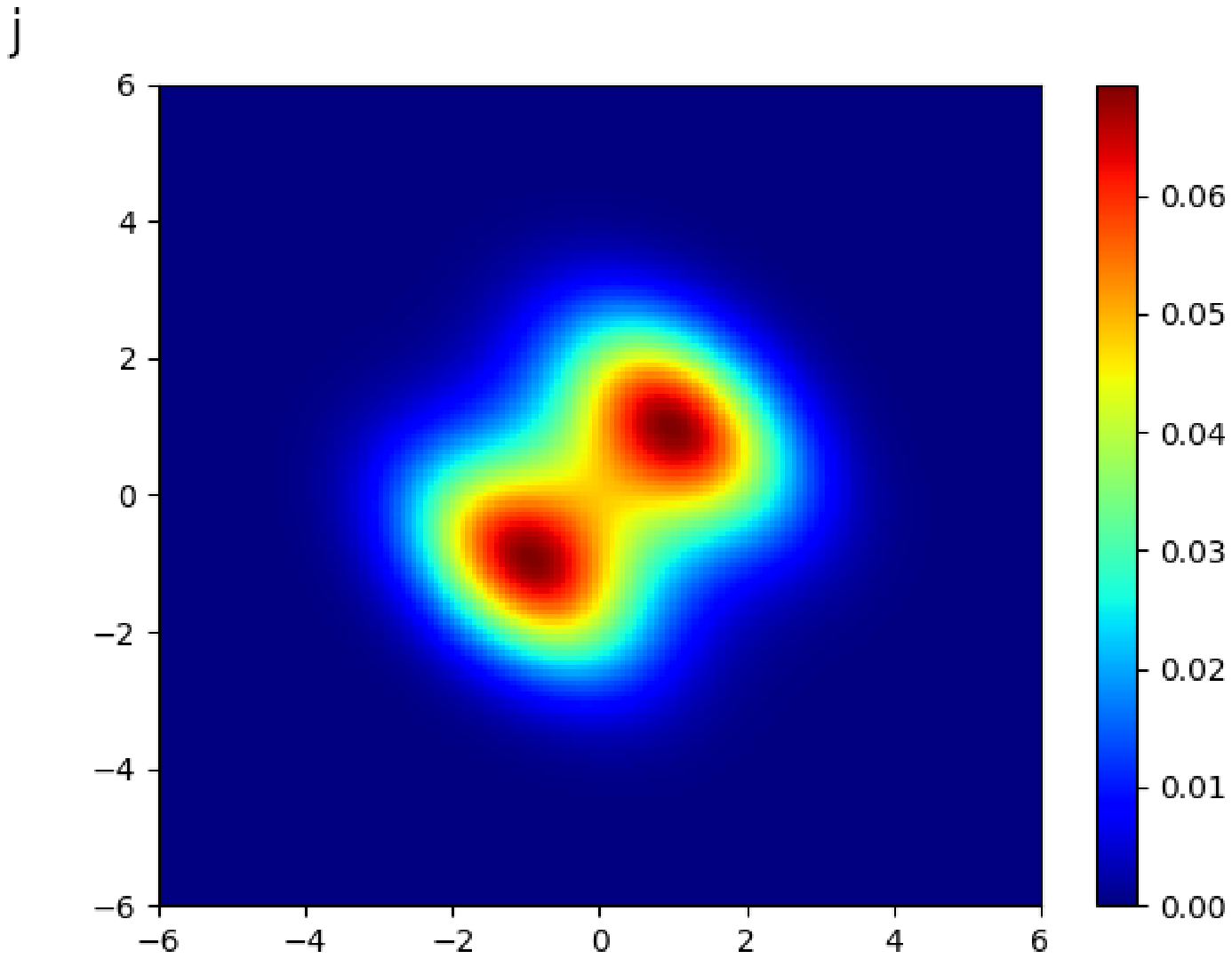}\\
\includegraphics[width=0.3\textwidth]{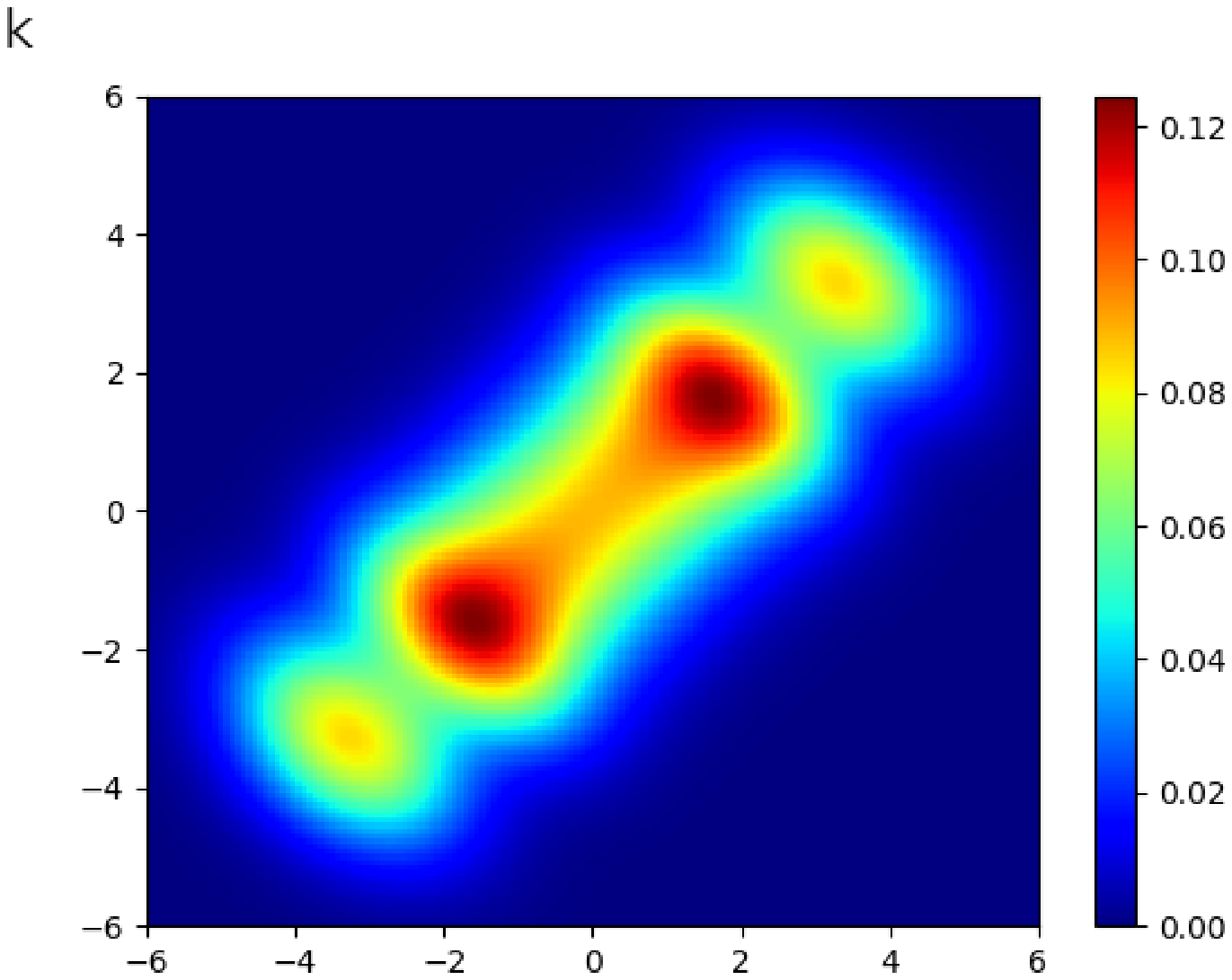}\quad
\includegraphics[width=0.3\textwidth]{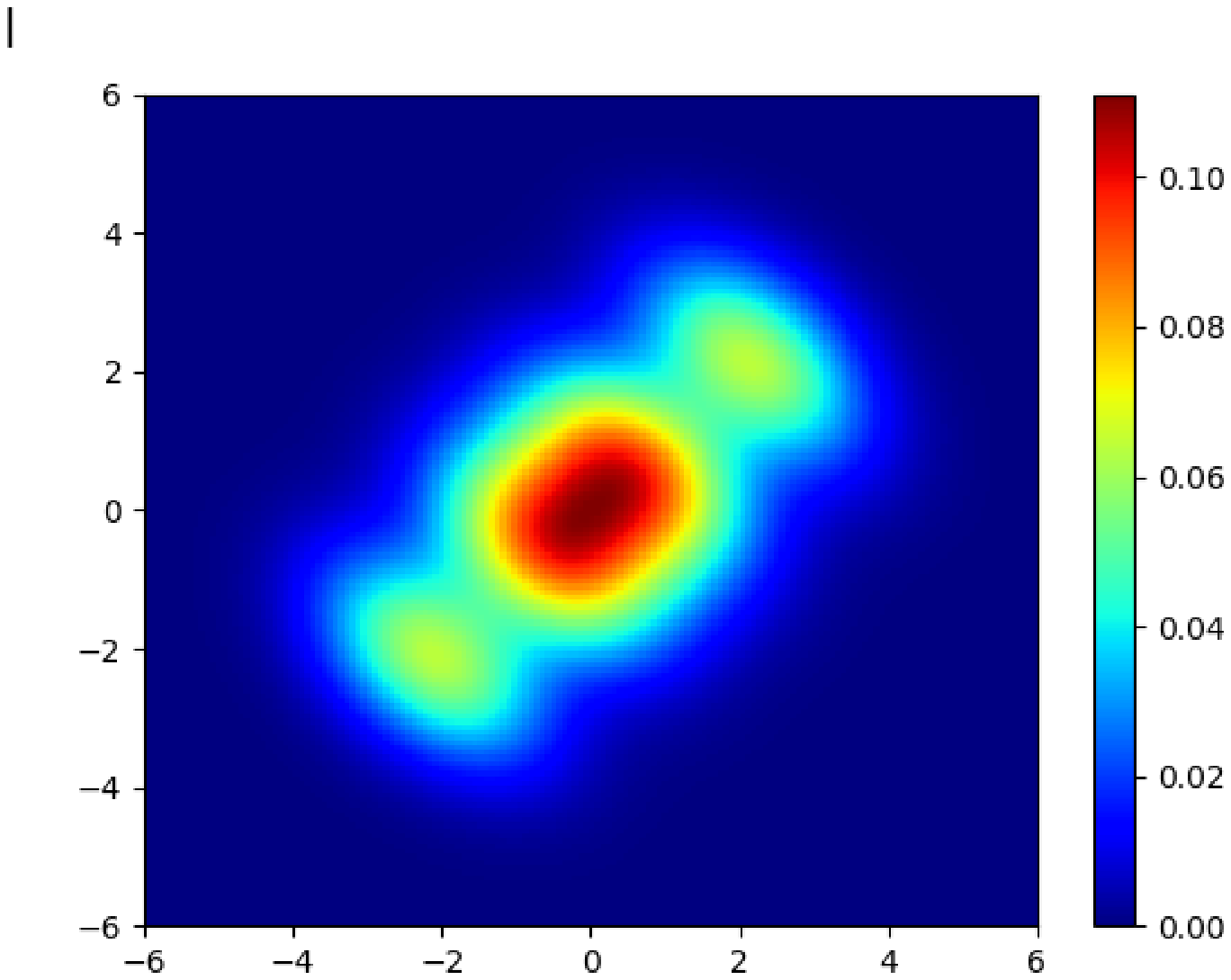}\quad
\includegraphics[width=0.3\textwidth]{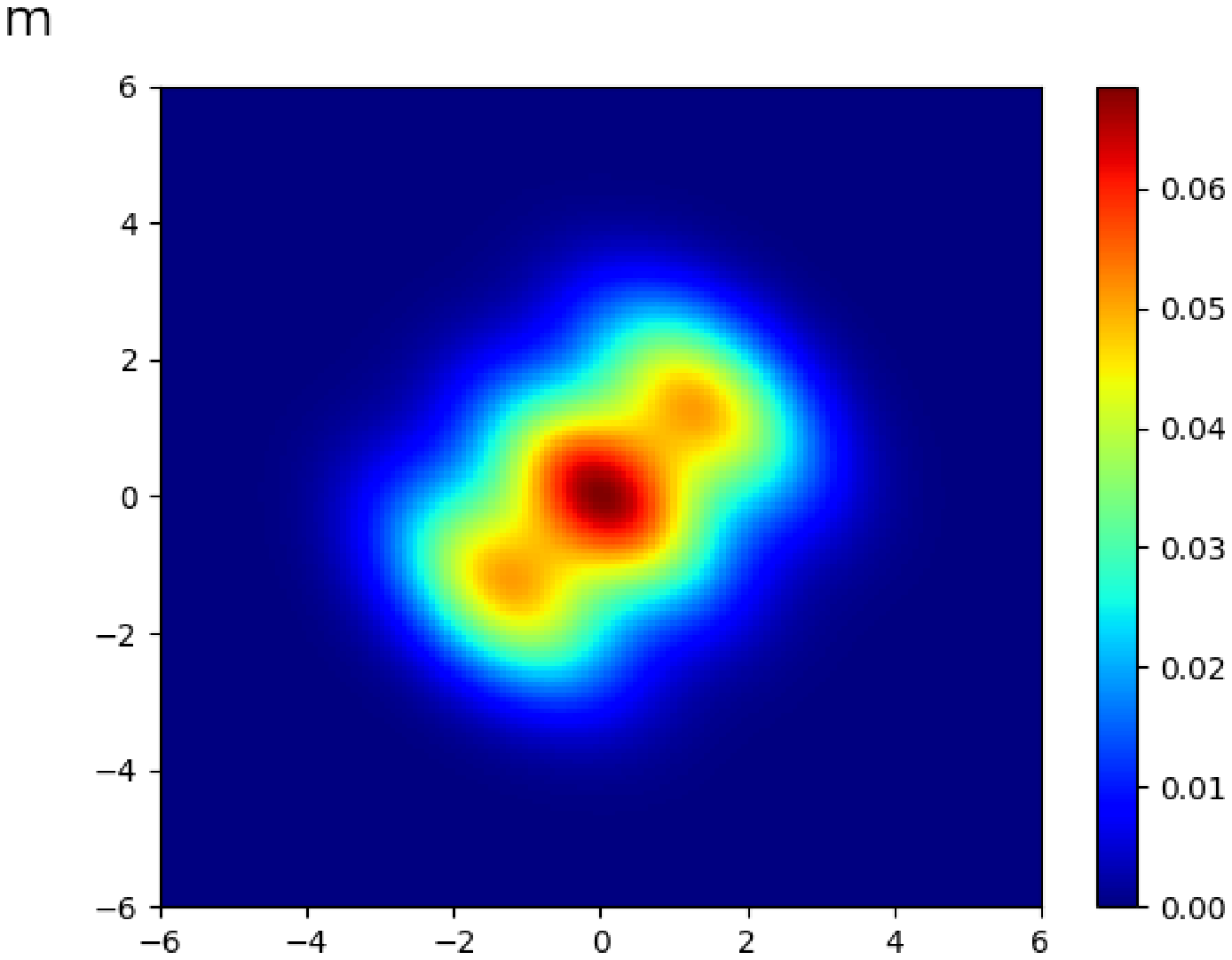}
\end{minipage}\\[1em]
\caption{Two-dimensional densities $\left(\phi(\mathbf{r})\psi_j(\mathbf{R})\right)^2$ of two electrons confined by a harmonic potential with
$\omega_0=1/3$ a.u. and spin $S=1$. 
First row: $j=0$, (a) $\lambda=0.01$, (b) $\lambda=0.5$, (c) $\lambda=2$.
Second row: $j=1$, (d) $\lambda=0.01$, (e) $\lambda=0.5$, (f) $\lambda=2$.
Third row: $j=2$, (g) $\lambda=0.01$, (h) $\lambda=0.5$, (j) $\lambda=2$.
Third row: $j=5$, (k) $\lambda=0.01$, (l) $\lambda=0.5$, (m) $\lambda=2$. 
}
\label{triplet}
\end{figure*}
\end{center}

In the  spin-triplet case in 2D we choose $\omega_0=1/3$ a.u., and  the energy of the relative motion is $\epsilon=1$ a.u. Fig. \ref{triplet}
shows the densities for this case. This system is more sensitive to the
choice of $\lambda$ and we use three different $\lambda$ values (0.01, 0.5, 2) to illustrate that. In this case, the spin function is symmetric, the  spatial part is antisymmetric ($m=\pm 1$ in Eq. (1.2) in Ref. \cite{Supp})  and the two peaks appear in the density 
plot for $j=0$ as shown in Figs. \ref{triplet}a, \ref{triplet}b and
\ref{triplet}c. By increasing $\lambda$ the $U$ oscillator squeezes the electrons closer  and the
separation between the two peaks is more visible (the density between the peaks being lower). There are three peaks for $j=1$ for $\lambda=0.01$ 
and $\lambda=0.5$, but as the $U$ confinement gets stronger the two
peak structure returns (Figs. \ref{triplet}d, \ref{triplet}e, and
\ref{triplet}f). The three-peak
structure can be a nontrivial consequence because, unlike the simple spherical structure in the singlet state, the relative motion function, in this case, is in an $m=1$ state and multiplied 
by $U$. 
If we neglect the Coulomb interaction, then the relative motion wave function is a ground state harmonic oscillator for the first electron
(one density peak) and the first excited harmonic oscillator state for the second electron (two peaks). These three peaks are magnified when the relative wave function is multiplied by the center
of mass wave function, which is now proportional to 
$H_1(\sqrt{\omega_U}\,U)=2\sqrt{\omega_U}\,U$. 

For higher $j$ states, the elongation caused by $H_j$ continues 
(see Figs. \ref{triplet}g and \ref{triplet}k), and the nodal structure 
of $H_j$ also contributes to the density structure. Overall it seems that the $\lambda=0.01$ case captures the general trend very well. For larger $\lambda$ values the same structures appear later as $j$ increases. For a given $j$, increasing $\lambda$ squeezes the elongation due to the $\omega_U$ confinement, as in the singlet case.

\begin{figure}
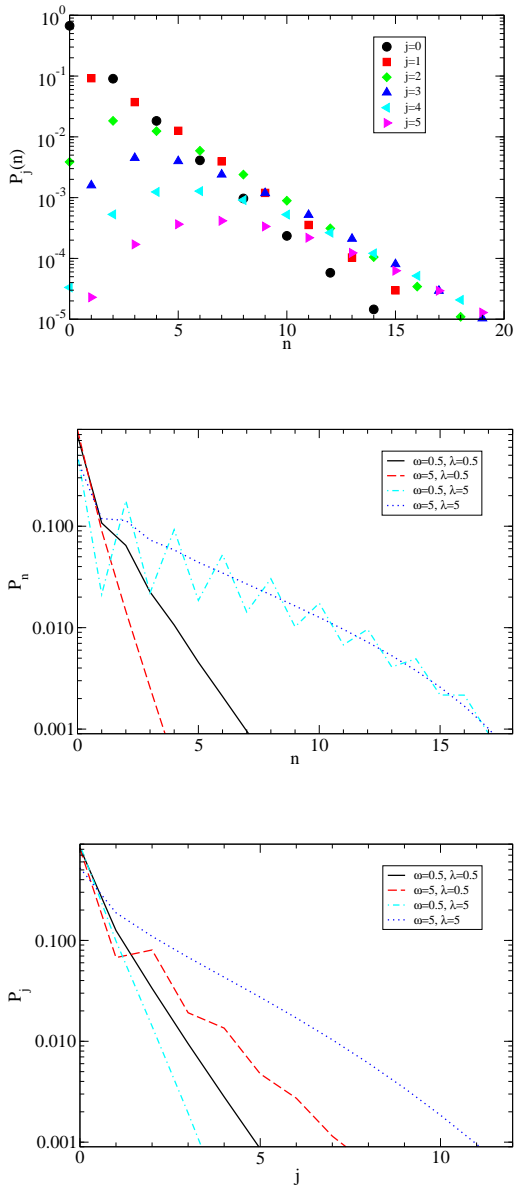

\begin{minipage}{0.5\textwidth}
\includegraphics[width=0.75\textwidth]{figure3a.eps} \\\vskip 1cm
\includegraphics[width=0.75\textwidth]{figure3b.eps}\\\vskip 1cm
\includegraphics[width=0.75\textwidth]{figure3c.eps}
\end{minipage}
\caption{Top: Probability of the occupation of a $jn$ subspace 
($\lambda=0.5$ a.u. and $\omega=0.5$ a.u.), $P_j(n)$.
Middle: $P_n=\sum_j P_n(j)$. Bottom :$P_j=\sum_n P_n(j)$. 
$\omega_0=1/3$ a.u. is used in the calculations.}
\label{prob}
\end{figure}

The total wave function will be a linear combination of 
the $\phi(\mathbf{r})\psi_j(\mathbf{R})|n\rangle$ components. The 
probability of a given component is given by
\begin{equation}
    P_j(n)=\vert c_{jn}^0 \vert^2
\end{equation}
and depends on $\omega_0, \omega$ and $\lambda$. An example for 
$P_j(n)$ is given in Fig. \ref{prob}a. First, we note that due to 
the structure in Eq. \eqref{matr}, the probabilities follow a checkerboard-like structure: odd photon numbers couple to odd $j$ and even photon numbers couple 
to even $j$. The probabilities decrease for large photon numbers. The low CM excitations, $j=0,1,3$, are the most dominant terms for low 
photon numbers. The probability of the higher CM 
excitations ($j=2,3,4,5$) first increases with the photon number, 
then reaches a maximum and starts to decrease. 

Fig. \ref{prob}b  shows the sum $P_n=\sum_j P_n(j)$. By increasing
$\lambda$, the higher photon spaces are coupled and the occupation of lower photon numbers increases. However, 
the effect of $\omega$ is more complicated. 
The coupling increases as $\sqrt{\omega}$ but with larger photon frequency the  photon harmonic oscillator states move higher in energy ($n\hbar \omega$)  and their occupation decreases. This latter
effect seems to be dominant for smaller $\lambda$. In Fig. \ref{prob}b, in case of $\lambda=0.5$ a.u., $P_n$ is the same for $n=0,1$ for $\omega=0.5$ a.u. and $\omega=5$ a.u., but for higher $n$, $P_n$ is much smaller for $\omega=5$ a.u. For higher $\lambda$ values this effect becomes less important. The oscillations (the even states have higher occupation than the odd states) in the case of $\omega$=0.5, $\lambda$=5 a.u. always appear when $\omega$ is much smaller than $\lambda$ and probably due to the checkerboard-like coupling.

Fig. \ref{prob}c  shows the sum $P_j=\sum_n P_n(j)$. By increasing $\omega$, the occupations of the 
low photon number states decrease and the occupations of the higher states increase. 
Increasing $\lambda$ increases $\omega_U$ and pushes the CM states higher, and those states do not
couple with the low $j$ sector, so increasing $\lambda$ decreases $P_j$.
The effect of $\lambda$ is similar to  $\omega$ in the previous case: $\lambda$ increases the coupling, 
but larger $\lambda$ means larger $\omega_u$ and the CM states are pushed higher. For low
$\omega$, increasing $\lambda$ relaxes the occupation, but for large
$\omega$, the coupling dominates and the $\lambda$ increases the occupation of the higher $n$ states. 

\begin{figure}
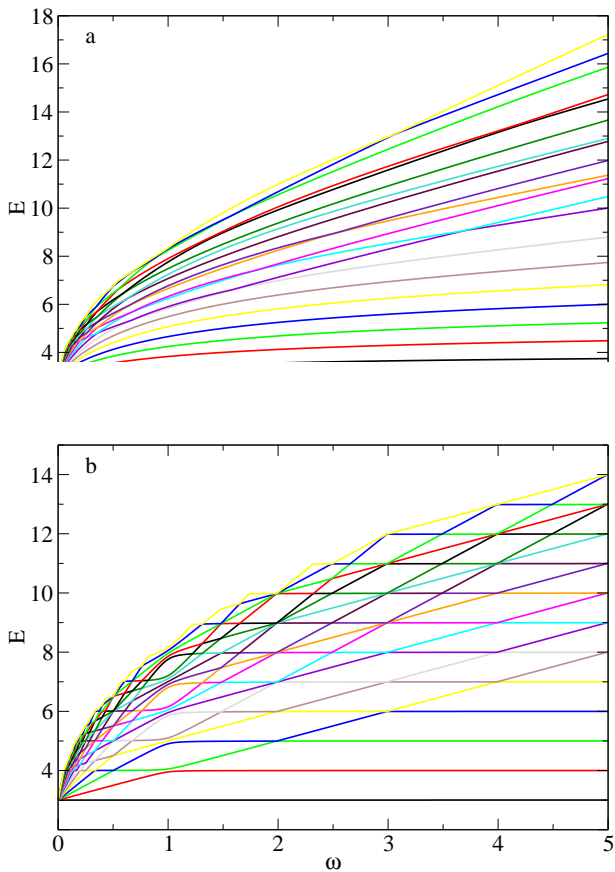

\includegraphics[width=0.45\textwidth]{figure4a.eps}\\
\includegraphics[width=0.45\textwidth]{figure4b.eps}
\caption{Energy levels as a function of $\omega$ for (a) $\lambda=\sqrt{\omega}$ and (b) $\lambda=\sqrt{\omega}/20$, $\omega_0$=1 a.u.
in both cases.}
\label{eomega}
\end{figure}

Fig. \ref{eomega} shows the energy of the 
singlet state as a function of photon frequency. 
The ground state energy is 3 a.u. in this case. Infinitely
many photon states and infinitely many CM states 
can couple to this state. Without coupling of the photons to the
center of mass, the  energy of the photon states increases
linearly with $\omega$ and the energy of the CM states 
increases linearly with $\omega_U$. Fig. \ref{eomega} shows the lowest
20 states with coupling, and we use $\lambda=\alpha \sqrt{\omega}$, which means that the coupling $g$ is proportional to $\omega$. For
$\alpha=1$ (Fig. \ref{eomega}a), some states (primarily photon states) move linearly up with
$\omega$ for small frequencies, while other states (primarily CM states) only slowly increase 
with $\omega$ and converge to a horizontal line. To magnify the behavior 
we redo the calculation with $\alpha=1/20$ (Fig. \ref{eomega}b). 
In this case, $\lambda$ increases much less while we increase $\omega$. $\omega_U$ barely changes 
while $\omega$ ascends from 0 to 5. The lowest state does not
change (it barely couples to photons $n>0$) and remains a horizontal line, which is just the lowest CM state. The second state increases with 
$\omega$, but then it reaches the first excited CM state
of energy $\hbar \omega_U\approx \hbar \omega_0$ and becomes a
horizontal line. The third state also increases until it reaches the
energy of the first excited state and continues on that line
until meeting the second state. To avoid crossing, it moves to the 
second excited state of the center of mass and so on. Of course, these
states do not lie exactly on horizontal lines but rise gradually with lambda. For much higher $\omega$, one can recover a similar picture to
Fig. \ref{eomega}a. 

To complete the energy spectrum of the system,
one has to include the excited states of the relative motion. As those
states are orthogonal, the complete spectrum can be obtained by shifting
the energy levels in Fig. \ref{eomega} by the energies of the excited
states.

\begin{figure}
\begin{minipage}{0.5\textwidth}
\includegraphics[width=0.45\textwidth]{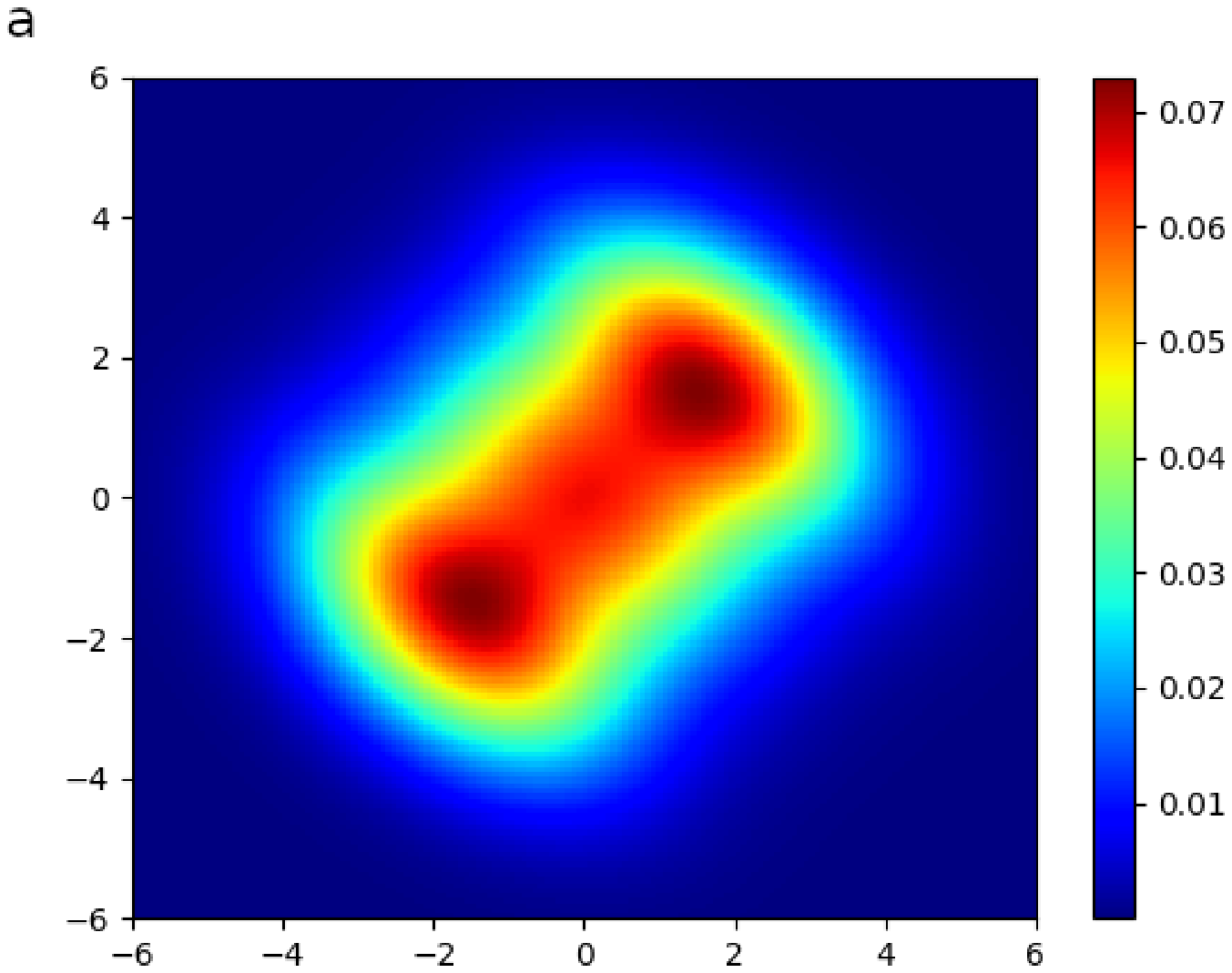}
\includegraphics[width=0.45\textwidth]{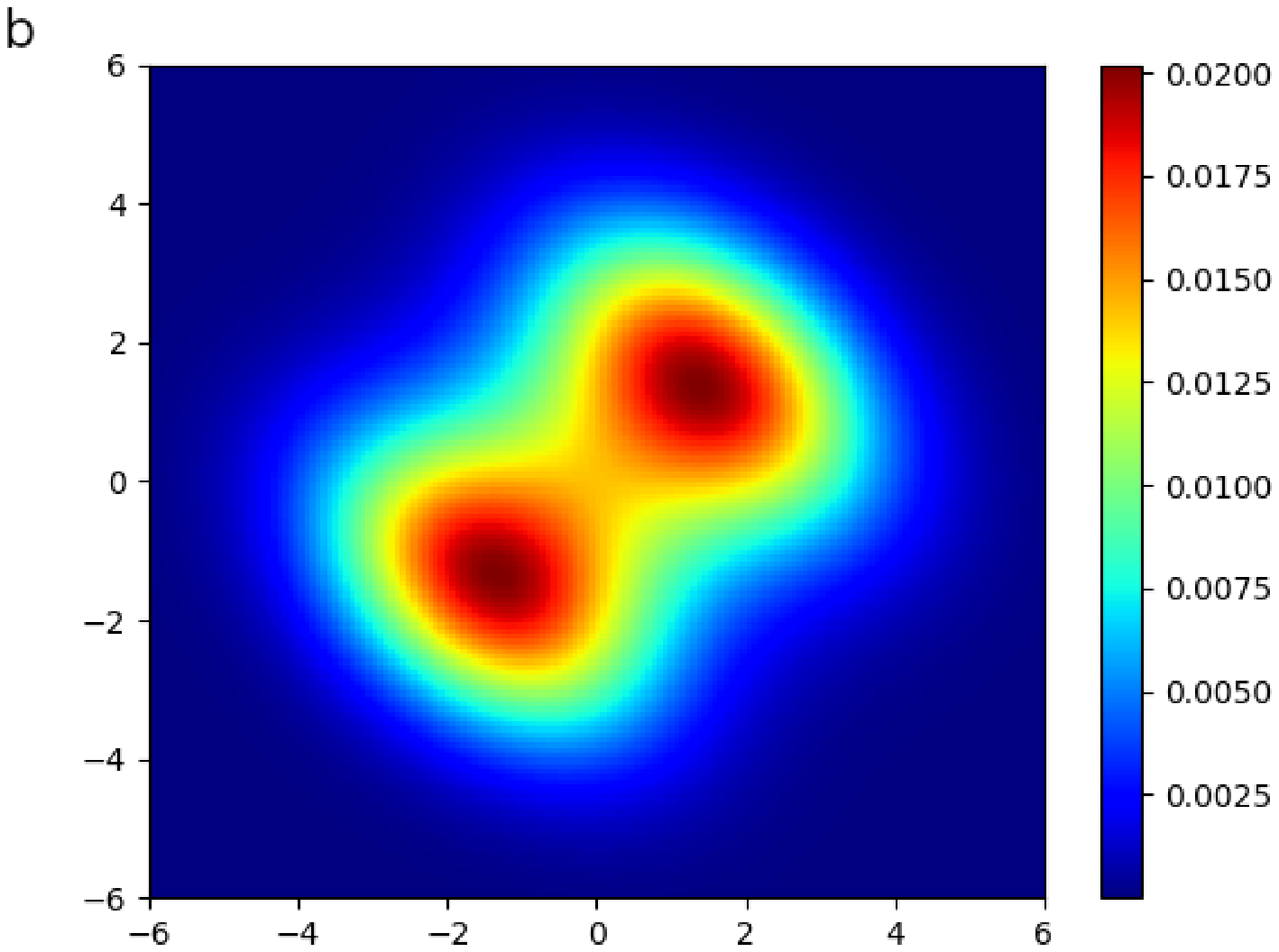}\\\quad
\includegraphics[width=0.45\textwidth]{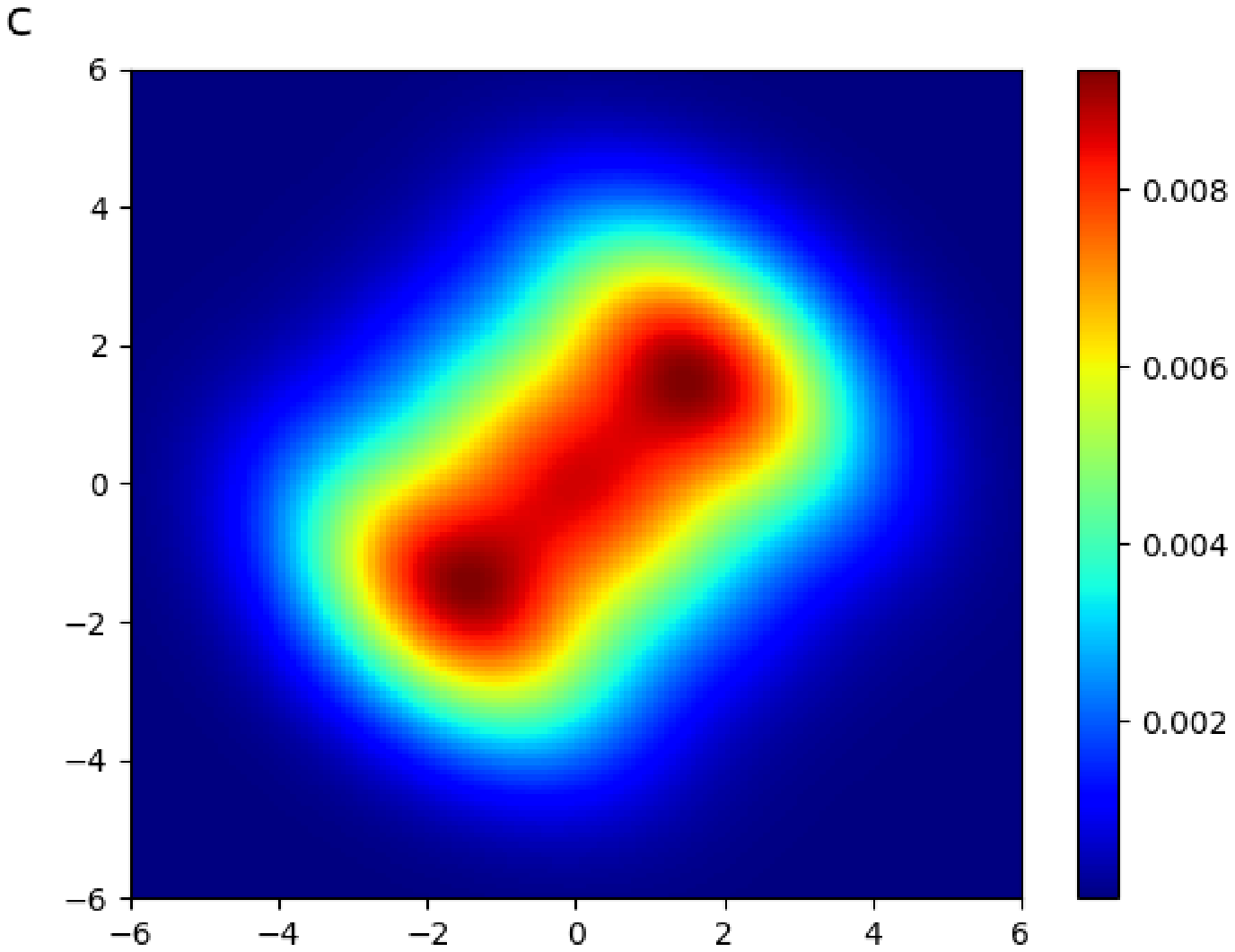}
\includegraphics[width=0.45\textwidth]{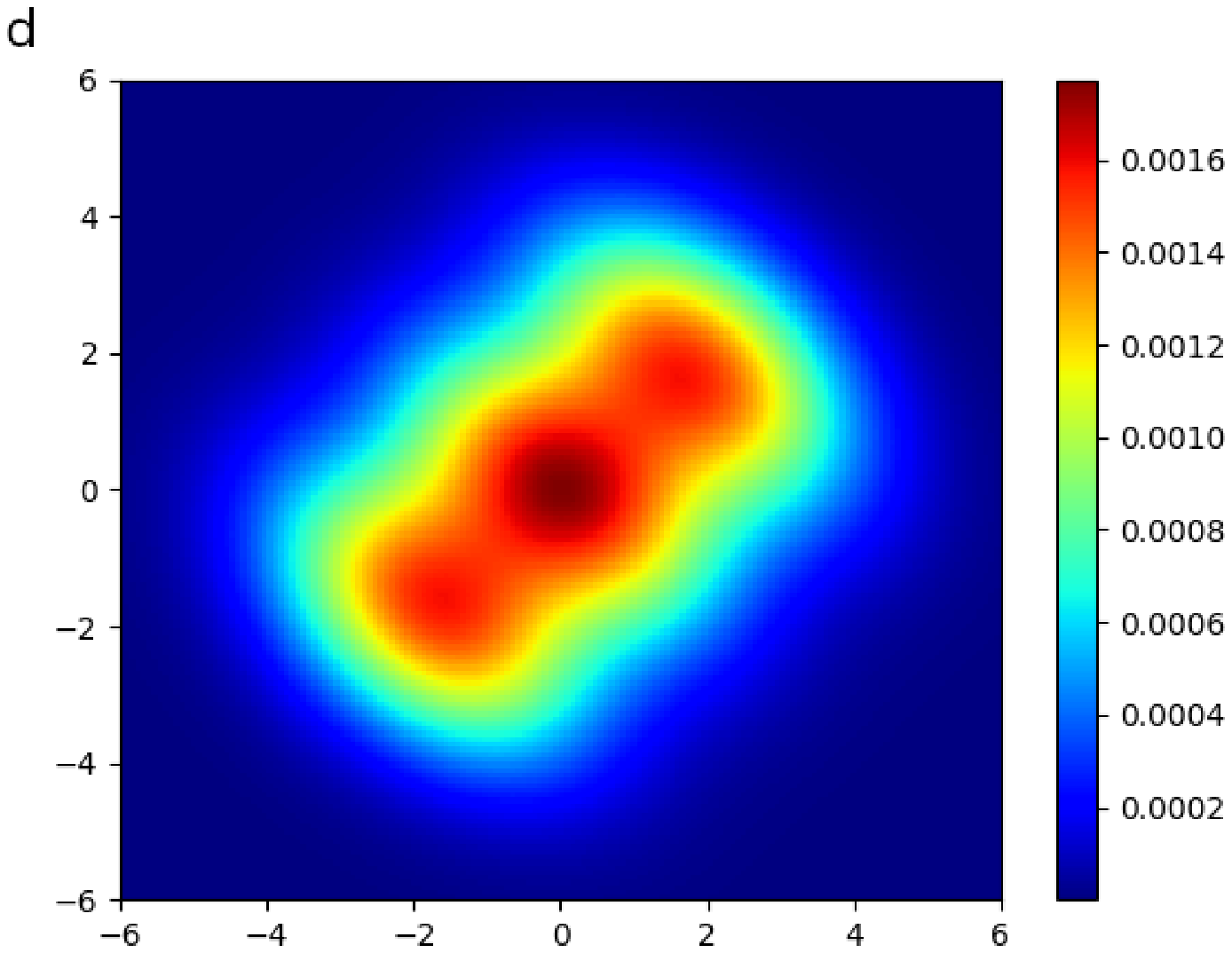}
\end{minipage}
\caption{Two-dimensional density $\Phi(\mathbf{r},\mathbf{R})^2$ 
for the $S=1$ case, (a) total density, (b) density in the $n=1$ space, (c) density in the $n=3$ space, (d) density in the $n=5$ space.
($\omega_0=0.18055$ a.u., $\omega=1$ a.u. and $\lambda=1$ a.u.).}
\label{triplet1}
\end{figure}

The wavefunction of the system will be a linear combination of wave functions shown in Figs. \ref{singlet} and \ref{triplet}, with coefficients defined in the second line of Eq. \eqref{trunc}. These coefficients depend on the values of $\omega$, $\omega_0$, and $\lambda$. For a single photon mode, the lowest states often 
dominate and it is hard to pick parameters that favor a single $j$ CM mode or higher $j$ values. In Fig. \ref{triplet1}a we present an example for the triplet case
where the square of the linear coefficients are 
0.55, 0.18, 0.10, 0.06, 0.04, 0.03 (j=0,...5), so a few $j\ne 0$
contribute to the density. Figs. \ref{triplet1}b, \ref{triplet1}c and
\ref{triplet1}d show the 
square of the wave function in the $n=1,3$ and 5 spaces. The $n$=0 density
is very similar to Fig. \ref{triplet}b. The squares of the linear coefficients 
in $n$ space are 0.49, 0.13, 0.11, 0.07, 0.05 ($n=0,...,5$). 
This example shows that the spatial wave functions in different photon
subspaces can be very different and quantum mechanical methods have to
look for accurate wave functions in different photon spaces.

\begin{figure}
\includegraphics[width=0.45\textwidth]{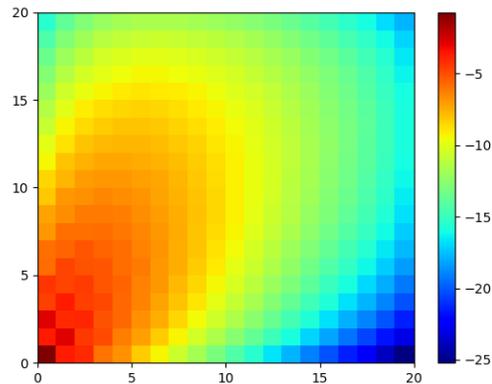}
\caption{Occupation numbers for the photon modes. The photon
frequencies are $\omega$  and $2\omega$. The coupling vectors
are $\boldsymbol{\lambda}$ and $-\boldsymbol{\lambda}$. The vertical axis is the photon number for $\omega$, the horizontal axis is the photon number for $2\omega$ ($\omega_0$=1 a.u., $\omega=1$ a.u., $\lambda=1$ a.u.).}
\label{2omega}
\end{figure}

\begin{figure}
\includegraphics[width=0.45\textwidth]{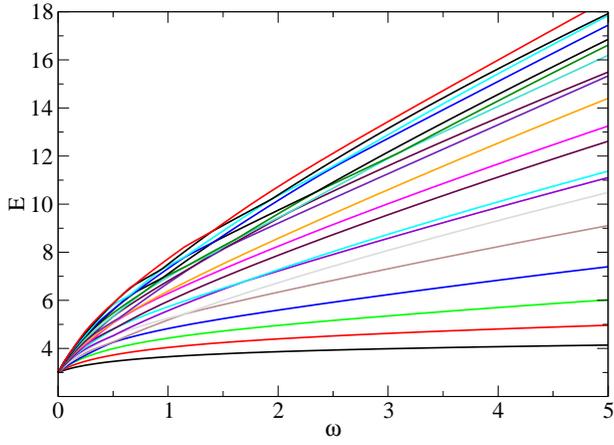}
\caption{Energy levels as a function of $\omega$ for the two photon modes. 
The photon frequencies are $\omega$ and $2\omega$, the coupling vectors
are $\boldsymbol{\lambda}$ and $-\boldsymbol{\lambda}$, with $\lambda=\sqrt{\omega}$ ($\omega_0$=1 a.u.)}
\label{e2omega}
\end{figure}

The calculation can be extended to many photon modes as it is shown
in Appendix \ref{nphoton}. Examples of two-photon mode calculations
are shown in Figs. \ref{2omega} and \ref{e2omega}. In particular, Fig. \ref{2omega} shows
the photon occupation numbers for the two-photon modes, $\omega$ and $2\omega$. The occupation probability tilts toward the $\omega$ axis, showing that the $\omega$ modes have higher probabilities than the $2\omega$ ones. Fig. \ref{e2omega} is the same calculation as is shown 
in Fig. \ref{eomega}a, but with two-photon modes. Overall, the two figures are very similar. The two-photon case reaches higher energies and there are more level crossings. This is because some of the states shown in Fig. \ref{e2omega} are $2\omega$ states and move higher faster.
Increasing the number of photon modes helps to reach higher $j$ states
and multiphoton modes might be a way to select higher $j$ states or single
out a desired $j$ value.


\section{Summary}
In a harmonically confined two-electron system, the light couples to the
dipole moment which is proportional to the CM coordinate. By separating the relative and center of mass motion, we have shown that the
coupled photon center of mass system can be solved by diagonalization and the relative motion part has analytical solutions for certain frequencies.

The coupling of the light to the center of mass coordinate leads to elongated wave functions. The symmetry axis of the electron density is determined by the
polarization direction. The density has several peaks depending on the center of mass excitation and the symmetry axis of the density is determined by the polarization direction. The competition between the confinement due to the coupling to light and the node structure of the center of mass excitation influences the location of the density peaks.

We have shown that the spatial wave functions belonging to different 
photon spaces are very different, and this means that quantum mechanical approaches solving coupled light-matter problems have to determine the
wave functions in each photon subspace, which might be a difficult task. 

The approach can be extended to many photon modes and the only limitation 
is the dimension of the Hamiltonian matrix. As this matrix is very sparse,
one can easily diagonalize it even for very large matrices. 

As there are only very few light-matter coupled systems with analytical
solutions, the present work might be useful to test and develop efficient 
approximations. 

A similar approach can be used for a larger electron number, but then the relative motion part has to be solved numerically. 
\clearpage
\appendix
\section{$N_p$ photon modes}
\label{nphoton}
Consider the same system as in Section \ref{like}, except that here $n_p$ photons are coupled. Hence, there are $N_p=2n_p$ photon modes involved.
\begin{equation}
|\vec{n}\rangle =|n_1,n_2,{\ldots},n_{N_p}\rangle.
\end{equation}

Define the vector Kronecker delta as
\begin{equation}
\delta_{\vec{n}\vec{m}}=\prod_{k=1}^{N_p} \delta_{n_k m_k},
\end{equation}

\begin{equation}
\delta_{\vec{n}\vec{m}}^l=\prod_{k=1,k\ne l}^{N_p} \delta_{n_k m_k}.
\end{equation}

It is straightforward to generalize Eq. \eqref{matr}
\begin{widetext}
\begin{equation}
    \langle\vec{m},\phi_i\vert H\vert \vec{n},\phi_j\rangle=
    \delta_{\vec{m}\vec{n}}\delta_{ij}(j+\frac{1}{2})\omega_U+
    \delta_{\vec{m}\vec{n}}\delta_{ij}\sum_{k=1}^{N_p}(n_k+\frac{1}{2})\omega_k+
    \sum_{k=1}^{N_p}\sqrt{\frac{2\omega_k}{\omega_U}}\,\lambda\,
    D_{n_km_k} D_{ij} \delta_{\vec{n}\vec{m}}^k.
\end{equation}
\end{widetext}

\section{Center-of-mass motion for many photons}
\label{lambda}
We assume $N_p=2n_p$ photon modes, and $\boldsymbol{\lambda_\alpha}$'s are not necessarily isotropic in the $x$,\,$y$-directions. Thus, the Hamiltonian becomes
\begin{eqnarray}
H&=&-\frac{1}{2} \nabla_{1}^{2}+\frac{1}{2} \omega_0^{2} \mathbf{r}_{1}^{2}-\frac{1}{2} \nabla_{2}^{2}+\frac{1}{2} \omega_0^{2} \mathbf{r}_{2}^{2}\nonumber \\ &+&\frac{q_1 q_2}{\left|\mathbf{r}_{1}-\mathbf{r}_{2}\right|} +{1\over 2}\sum_{\alpha=0}^{N_p} (q_1\boldsymbol{\lambda}_\alpha\cdot\mathbf{r}_1+
q_2\boldsymbol{\lambda}_\alpha\cdot\mathbf{r}_2)^2.
\end{eqnarray}

Still imposing $q_1=q_2$, the radial part remains unchanged and can be
solved by Ref\cite{taut,taut2,Supp}. Now we solve the CM part. Eq. \eqref{equ2} becomes 
\begin{equation}
\left[-\frac{1}{2} \nabla_{\mathbf{R}}^{2}+\frac{1}{2} \omega_{\mathbf{R}}^{2} \mathbf{R}^{2}+4\sum_{\alpha=0}^{N_p}(\boldsymbol{\lambda}_\alpha\cdot\mathbf{R})^2
\right] \xi(\mathbf{R})=\eta^{\prime} \xi(\mathbf{R}).
\end{equation}
Suppose $\boldsymbol{\lambda}_\alpha=(\lambda_{\alpha 1},\lambda_{\alpha 2},0)$. Further define
\begin{equation}
\begin{aligned}
    \tilde{\lambda}_1&=\sum_{\alpha=0}^{N_p}\lambda_{\alpha 1}\strut^2,\\
    \tilde{\lambda}_2&=\sum_{\alpha=0}^{N_p}\lambda_{\alpha 2}\strut^2,\\
    \tilde{\lambda}_{12}&=\displaystyle\sum_{\alpha=0}^{N_p}\lambda_{\alpha 1}\lambda_{\alpha 2},
\end{aligned}    
\end{equation}
Eq. \eqref{equ3} and \eqref{equ4} now read

\begin{equation}
H_{\mathbf{R}}=
-{1\over 2}{\partial^2 \over \partial X^2}
-{1\over 2}{\partial^2 \over \partial Y^2}
+{1\over 2}\omega_X^2 X^2+{1\over 2}\omega_Y^2 Y^2 +{1\over 2}\omega_{XY} XY,
\end{equation}
where 
\begin{equation}
\begin{aligned}
    &\omega_X^2=\omega_{\mathbf{R}}^2+8{\tilde{\lambda}_1},\\
    &\omega_Y^2=\omega_{\mathbf{R}}^2+8{\tilde{\lambda}_2},\\
    &\omega_{XY}=16\tilde{\lambda}_{12}.
\end{aligned}
\end{equation}

This linearly coupled Hamiltonian can be easily decoupled with the following unitary transformation
\begin{equation}
\begin{aligned}
    U=\frac{1}{(1-ab)^{1/2}}(X+aY),\\
    V=\frac{1}{(1-ab)^{1/2}}(bX+Y),
\end{aligned}
\end{equation}
where
\begin{equation}
\begin{aligned}
    a=\frac{(\tilde{\lambda}_1-\tilde{\lambda}_2)+\sqrt{\,(\tilde{\lambda}_1-\tilde{\lambda}_2)\,\strut^2+4\tilde{\lambda}_{12}\strut^2}}{2\tilde{\lambda}_{12}},\\
    b=-\frac{(\tilde{\lambda}_1-\tilde{\lambda}_2)+\sqrt{\,(\tilde{\lambda}_1-\tilde{\lambda}_2)\,\strut^2+4\tilde{\lambda}_{12}\strut^2}}{2\tilde{\lambda}_{12}}.
\end{aligned}    
\end{equation}\\
In this case, the decoupled Hamiltonian reads
\begin{equation}
    H_R(U,V)=-\frac{1}{2}\frac{\partial^2}{\partial U^2}-\frac{1}{2}\frac{\partial^2}{\partial V^2}+\frac{1}{2}\omega_U^2 U^2+\frac{1}{2}\omega_V^2 V^2,
\end{equation}
where
\begin{equation}
\begin{aligned}
    \omega_U=\sqrt{\omega_R^2+4(\tilde{\lambda}_1+\tilde{\lambda}_2)+4\sqrt{(\tilde{\lambda}_1-\tilde{\lambda}_2)\strut^2+4\tilde{\lambda}_{12}\strut^2}},\\
    \omega_V=\sqrt{\omega_R^2+4(\tilde{\lambda}_1+\tilde{\lambda}_2)-4\sqrt{(\tilde{\lambda}_1-\tilde{\lambda}_2)\strut^2+4\tilde{\lambda}_{12}\strut^2}}.
\end{aligned}
\end{equation}

Same as in Eq. \eqref{harm}, this is just the Hamiltonian for two non-interacting harmonic oscillators. Hence, the energies for the CM part are
\begin{equation}
    \eta=\frac{1}{2}\eta^\prime=\frac{1}{2}(n\strut_U+\frac{1}{2})\omega_U+\frac{1}{2}(n\strut_V+\frac{1}{2})\omega_V\,\text{, } n\strut_U,\ n\strut_V=0,1,2...
\end{equation}
and the ground state energy is
\begin{equation}
    \eta_0=\frac{1}{4}(\omega_U+\omega_V).
\end{equation}

Finally, the corresponding wave function is just the product of that of the two independent harmonic oscillators.

\begin{acknowledgments}
This work has been supported by the National Science
Foundation (NSF) under Grant No. IRES 1826917.
\end{acknowledgments}


\clearpage
\begin{thebibliography}{53}
\expandafter\ifx\csname natexlab\endcsname\relax\def\natexlab#1{#1}\fi
\expandafter\ifx\csname bibnamefont\endcsname\relax
  \def\bibnamefont#1{#1}\fi
\expandafter\ifx\csname bibfnamefont\endcsname\relax
  \def\bibfnamefont#1{#1}\fi
\expandafter\ifx\csname citenamefont\endcsname\relax
  \def\citenamefont#1{#1}\fi
\expandafter\ifx\csname url\endcsname\relax
  \def\url#1{\texttt{#1}}\fi
\expandafter\ifx\csname urlprefix\endcsname\relax\def\urlprefix{URL }\fi
\providecommand{\bibinfo}[2]{#2}
\providecommand{\eprint}[2][]{\url{#2}}

\bibitem[{\citenamefont{Buchholz et~al.}(2019)\citenamefont{Buchholz,
  Theophilou, Nielsen, Ruggenthaler, and
  Rubio}}]{doi:10.1021/acsphotonics.9b00648}
\bibinfo{author}{\bibfnamefont{F.}~\bibnamefont{Buchholz}},
  \bibinfo{author}{\bibfnamefont{I.}~\bibnamefont{Theophilou}},
  \bibinfo{author}{\bibfnamefont{S.~E.~B.} \bibnamefont{Nielsen}},
  \bibinfo{author}{\bibfnamefont{M.}~\bibnamefont{Ruggenthaler}},
  \bibnamefont{and} \bibinfo{author}{\bibfnamefont{A.}~\bibnamefont{Rubio}},
  \bibinfo{journal}{ACS Photonics} \textbf{\bibinfo{volume}{6}},
  \bibinfo{pages}{2694} (\bibinfo{year}{2019}).

\bibitem[{\citenamefont{Sch{\"a}fer et~al.}(2019)\citenamefont{Sch{\"a}fer,
  Ruggenthaler, Appel, and Rubio}}]{Schafer4883}
\bibinfo{author}{\bibfnamefont{C.}~\bibnamefont{Sch{\"a}fer}},
  \bibinfo{author}{\bibfnamefont{M.}~\bibnamefont{Ruggenthaler}},
  \bibinfo{author}{\bibfnamefont{H.}~\bibnamefont{Appel}}, \bibnamefont{and}
  \bibinfo{author}{\bibfnamefont{A.}~\bibnamefont{Rubio}},
  \bibinfo{journal}{Proceedings of the National Academy of Sciences}
  \textbf{\bibinfo{volume}{116}}, \bibinfo{pages}{4883} (\bibinfo{year}{2019}),
  ISSN \bibinfo{issn}{0027-8424},
  \eprint{https://www.pnas.org/content/116/11/4883.full.pdf},
  \urlprefix\url{https://www.pnas.org/content/116/11/4883}.

\bibitem[{\citenamefont{Ruggenthaler et~al.}(2018)\citenamefont{Ruggenthaler,
  Tancogne-Dejean, Flick, Appel, and Rubio}}]{Ruggenthaler2018}
\bibinfo{author}{\bibfnamefont{M.}~\bibnamefont{Ruggenthaler}},
  \bibinfo{author}{\bibfnamefont{N.}~\bibnamefont{Tancogne-Dejean}},
  \bibinfo{author}{\bibfnamefont{J.}~\bibnamefont{Flick}},
  \bibinfo{author}{\bibfnamefont{H.}~\bibnamefont{Appel}}, \bibnamefont{and}
  \bibinfo{author}{\bibfnamefont{A.}~\bibnamefont{Rubio}},
  \bibinfo{journal}{Nature Reviews Chemistry} \textbf{\bibinfo{volume}{2}},
  \bibinfo{pages}{0118} (\bibinfo{year}{2018}), ISSN \bibinfo{issn}{2397-3358},
  \urlprefix\url{https://doi.org/10.1038/s41570-018-0118}.

\bibitem[{\citenamefont{Flick et~al.}(2015)\citenamefont{Flick, Ruggenthaler,
  Appel, and Rubio}}]{Flick15285}
\bibinfo{author}{\bibfnamefont{J.}~\bibnamefont{Flick}},
  \bibinfo{author}{\bibfnamefont{M.}~\bibnamefont{Ruggenthaler}},
  \bibinfo{author}{\bibfnamefont{H.}~\bibnamefont{Appel}}, \bibnamefont{and}
  \bibinfo{author}{\bibfnamefont{A.}~\bibnamefont{Rubio}},
  \bibinfo{journal}{Proceedings of the National Academy of Sciences}
  \textbf{\bibinfo{volume}{112}}, \bibinfo{pages}{15285}
  (\bibinfo{year}{2015}), ISSN \bibinfo{issn}{0027-8424},
  \eprint{https://www.pnas.org/content/112/50/15285.full.pdf},
  \urlprefix\url{https://www.pnas.org/content/112/50/15285}.

\bibitem[{\citenamefont{Flick et~al.}(2017)\citenamefont{Flick, Ruggenthaler,
  Appel, and Rubio}}]{Flick3026}
\bibinfo{author}{\bibfnamefont{J.}~\bibnamefont{Flick}},
  \bibinfo{author}{\bibfnamefont{M.}~\bibnamefont{Ruggenthaler}},
  \bibinfo{author}{\bibfnamefont{H.}~\bibnamefont{Appel}}, \bibnamefont{and}
  \bibinfo{author}{\bibfnamefont{A.}~\bibnamefont{Rubio}},
  \bibinfo{journal}{Proceedings of the National Academy of Sciences}
  \textbf{\bibinfo{volume}{114}}, \bibinfo{pages}{3026} (\bibinfo{year}{2017}),
  ISSN \bibinfo{issn}{0027-8424},
  \eprint{https://www.pnas.org/content/114/12/3026.full.pdf},
  \urlprefix\url{https://www.pnas.org/content/114/12/3026}.

\bibitem[{\citenamefont{Rokaj et~al.}(2018)\citenamefont{Rokaj, Welakuh,
  Ruggenthaler, and Rubio}}]{Rokaj_2018}
\bibinfo{author}{\bibfnamefont{V.}~\bibnamefont{Rokaj}},
  \bibinfo{author}{\bibfnamefont{D.~M.} \bibnamefont{Welakuh}},
  \bibinfo{author}{\bibfnamefont{M.}~\bibnamefont{Ruggenthaler}},
  \bibnamefont{and} \bibinfo{author}{\bibfnamefont{A.}~\bibnamefont{Rubio}},
  \bibinfo{journal}{Journal of Physics B: Atomic, Molecular and Optical
  Physics} \textbf{\bibinfo{volume}{51}}, \bibinfo{pages}{034005}
  (\bibinfo{year}{2018}),
  \urlprefix\url{https://doi.org/10.1088/1361-6455/aa9c99}.

\bibitem[{\citenamefont{Rivera et~al.}(2019)\citenamefont{Rivera, Flick, and
  Narang}}]{PhysRevLett.122.193603}
\bibinfo{author}{\bibfnamefont{N.}~\bibnamefont{Rivera}},
  \bibinfo{author}{\bibfnamefont{J.}~\bibnamefont{Flick}}, \bibnamefont{and}
  \bibinfo{author}{\bibfnamefont{P.}~\bibnamefont{Narang}},
  \bibinfo{journal}{Phys. Rev. Lett.} \textbf{\bibinfo{volume}{122}},
  \bibinfo{pages}{193603} (\bibinfo{year}{2019}),
  \urlprefix\url{https://link.aps.org/doi/10.1103/PhysRevLett.122.193603}.

\bibitem[{\citenamefont{Flick and Narang}(2018)}]{PhysRevLett.121.113002}
\bibinfo{author}{\bibfnamefont{J.}~\bibnamefont{Flick}} \bibnamefont{and}
  \bibinfo{author}{\bibfnamefont{P.}~\bibnamefont{Narang}},
  \bibinfo{journal}{Phys. Rev. Lett.} \textbf{\bibinfo{volume}{121}},
  \bibinfo{pages}{113002} (\bibinfo{year}{2018}),
  \urlprefix\url{https://link.aps.org/doi/10.1103/PhysRevLett.121.113002}.

\bibitem[{\citenamefont{Hoffmann et~al.}(2020)\citenamefont{Hoffmann, Lacombe,
  Rubio, and Maitra}}]{doi:10.1063/5.0012723}
\bibinfo{author}{\bibfnamefont{N.~M.} \bibnamefont{Hoffmann}},
  \bibinfo{author}{\bibfnamefont{L.}~\bibnamefont{Lacombe}},
  \bibinfo{author}{\bibfnamefont{A.}~\bibnamefont{Rubio}}, \bibnamefont{and}
  \bibinfo{author}{\bibfnamefont{N.~T.} \bibnamefont{Maitra}},
  \bibinfo{journal}{The Journal of Chemical Physics}
  \textbf{\bibinfo{volume}{153}}, \bibinfo{pages}{104103}
  (\bibinfo{year}{2020}).

\bibitem[{\citenamefont{Tokatly}(2018)}]{PhysRevB.98.235123}
\bibinfo{author}{\bibfnamefont{I.~V.} \bibnamefont{Tokatly}},
  \bibinfo{journal}{Phys. Rev. B} \textbf{\bibinfo{volume}{98}},
  \bibinfo{pages}{235123} (\bibinfo{year}{2018}),
  \urlprefix\url{https://link.aps.org/doi/10.1103/PhysRevB.98.235123}.

\bibitem[{\citenamefont{Galego et~al.}(2017)\citenamefont{Galego, Garcia-Vidal,
  and Feist}}]{PhysRevLett.119.136001}
\bibinfo{author}{\bibfnamefont{J.}~\bibnamefont{Galego}},
  \bibinfo{author}{\bibfnamefont{F.~J.} \bibnamefont{Garcia-Vidal}},
  \bibnamefont{and} \bibinfo{author}{\bibfnamefont{J.}~\bibnamefont{Feist}},
  \bibinfo{journal}{Phys. Rev. Lett.} \textbf{\bibinfo{volume}{119}},
  \bibinfo{pages}{136001} (\bibinfo{year}{2017}),
  \urlprefix\url{https://link.aps.org/doi/10.1103/PhysRevLett.119.136001}.

\bibitem[{\citenamefont{Mandal et~al.}(2020{\natexlab{a}})\citenamefont{Mandal,
  Montillo~Vega, and Huo}}]{Mandal}
\bibinfo{author}{\bibfnamefont{A.}~\bibnamefont{Mandal}},
  \bibinfo{author}{\bibfnamefont{S.}~\bibnamefont{Montillo~Vega}},
  \bibnamefont{and} \bibinfo{author}{\bibfnamefont{P.}~\bibnamefont{Huo}},
  \bibinfo{journal}{The Journal of Physical Chemistry Letters}
  \textbf{\bibinfo{volume}{11}}, \bibinfo{pages}{9215}
  (\bibinfo{year}{2020}{\natexlab{a}}), \bibinfo{note}{pMID: 32991814}.

\bibitem[{\citenamefont{Cederbaum and Kuleff}(2021)}]{Cederbaum2021}
\bibinfo{author}{\bibfnamefont{L.~S.} \bibnamefont{Cederbaum}}
  \bibnamefont{and} \bibinfo{author}{\bibfnamefont{A.~I.}
  \bibnamefont{Kuleff}}, \bibinfo{journal}{Nature Communications}
  \textbf{\bibinfo{volume}{12}}, \bibinfo{pages}{4083} (\bibinfo{year}{2021}),
  ISSN \bibinfo{issn}{2041-1723},
  \urlprefix\url{https://doi.org/10.1038/s41467-021-24221-6}.

\bibitem[{\citenamefont{Szidarovszky et~al.}(2018)\citenamefont{Szidarovszky,
  Hal\'asz, Cs\'asz\'ar, Cederbaum, and
  Vib\'ok}}]{doi:10.1021/acs.jpclett.8b02609}
\bibinfo{author}{\bibfnamefont{T.}~\bibnamefont{Szidarovszky}},
  \bibinfo{author}{\bibfnamefont{G.~J.} \bibnamefont{Hal\'asz}},
  \bibinfo{author}{\bibfnamefont{A.~G.} \bibnamefont{Cs\'asz\'ar}},
  \bibinfo{author}{\bibfnamefont{L.~S.} \bibnamefont{Cederbaum}},
  \bibnamefont{and} \bibinfo{author}{\bibfnamefont{A.}~\bibnamefont{Vib\'ok}},
  \bibinfo{journal}{The Journal of Physical Chemistry Letters}
  \textbf{\bibinfo{volume}{9}}, \bibinfo{pages}{6215} (\bibinfo{year}{2018}).

\bibitem[{\citenamefont{Ashida et~al.}(2021)\citenamefont{Ashida, \ifmmode
  \dot{I}\else \.{I}\fi{}mamo\ifmmode~\breve{g}\else \u{g}\fi{}lu, and
  Demler}}]{PhysRevLett.126.153603}
\bibinfo{author}{\bibfnamefont{Y.}~\bibnamefont{Ashida}},
  \bibinfo{author}{\bibfnamefont{A.~m.~c.} \bibnamefont{\ifmmode \dot{I}\else
  \.{I}\fi{}mamo\ifmmode~\breve{g}\else \u{g}\fi{}lu}}, \bibnamefont{and}
  \bibinfo{author}{\bibfnamefont{E.}~\bibnamefont{Demler}},
  \bibinfo{journal}{Phys. Rev. Lett.} \textbf{\bibinfo{volume}{126}},
  \bibinfo{pages}{153603} (\bibinfo{year}{2021}),
  \urlprefix\url{https://link.aps.org/doi/10.1103/PhysRevLett.126.153603}.

\bibitem[{\citenamefont{Rivera and Kaminer}(2020)}]{Rivera2020}
\bibinfo{author}{\bibfnamefont{N.}~\bibnamefont{Rivera}} \bibnamefont{and}
  \bibinfo{author}{\bibfnamefont{I.}~\bibnamefont{Kaminer}},
  \bibinfo{journal}{Nature Reviews Physics} \textbf{\bibinfo{volume}{2}},
  \bibinfo{pages}{538} (\bibinfo{year}{2020}), ISSN \bibinfo{issn}{2522-5820},
  \urlprefix\url{https://doi.org/10.1038/s42254-020-0224-2}.

\bibitem[{\citenamefont{Le~Boité}(2020)}]{https://doi.org/10.1002/qute.201900140}
\bibinfo{author}{\bibfnamefont{A.}~\bibnamefont{Le~Boité}},
  \bibinfo{journal}{Advanced Quantum Technologies}
  \textbf{\bibinfo{volume}{3}}, \bibinfo{pages}{1900140}
  (\bibinfo{year}{2020}),
  \eprint{https://onlinelibrary.wiley.com/doi/pdf/10.1002/qute.201900140},
  \urlprefix\url{https://onlinelibrary.wiley.com/doi/abs/10.1002/qute.201900140}.

\bibitem[{\citenamefont{Garcia-Vidal et~al.}(2021)\citenamefont{Garcia-Vidal,
  Ciuti, and Ebbesen}}]{Garcia-Vidaleabd0336}
\bibinfo{author}{\bibfnamefont{F.~J.} \bibnamefont{Garcia-Vidal}},
  \bibinfo{author}{\bibfnamefont{C.}~\bibnamefont{Ciuti}}, \bibnamefont{and}
  \bibinfo{author}{\bibfnamefont{T.~W.} \bibnamefont{Ebbesen}},
  \bibinfo{journal}{Science} \textbf{\bibinfo{volume}{373}}
  (\bibinfo{year}{2021}), ISSN \bibinfo{issn}{0036-8075},
  \eprint{https://science.sciencemag.org/content/373/6551/eabd0336.full.pdf},
  \urlprefix\url{https://science.sciencemag.org/content/373/6551/eabd0336}.

\bibitem[{\citenamefont{Taut}(1993)}]{taut}
\bibinfo{author}{\bibfnamefont{M.}~\bibnamefont{Taut}}, \bibinfo{journal}{Phys.
  Rev. A} \textbf{\bibinfo{volume}{48}}, \bibinfo{pages}{3561}
  (\bibinfo{year}{1993}),
  \urlprefix\url{https://link.aps.org/doi/10.1103/PhysRevA.48.3561}.

\bibitem[{\citenamefont{Taut}(1994)}]{taut2}
\bibinfo{author}{\bibfnamefont{M.}~\bibnamefont{Taut}},
  \bibinfo{journal}{Journal of Physics A: Mathematical and General}
  \textbf{\bibinfo{volume}{27}}, \bibinfo{pages}{1045} (\bibinfo{year}{1994}),
  \urlprefix\url{https://iopscience.iop.org/article/10.1088/0305-4470/27/3/040/pdf}.

\bibitem[{\citenamefont{Taut}(1995)}]{taut3}
\bibinfo{author}{\bibfnamefont{M.}~\bibnamefont{Taut}},
  \bibinfo{journal}{Journal of Physics A: Mathematical and General}
  \textbf{\bibinfo{volume}{28}}, \bibinfo{pages}{2081} (\bibinfo{year}{1995}),
  \urlprefix\url{https://iopscience.iop.org/article/10.1088/0305-4470/28/7/026/pdf}.

\bibitem[{\citenamefont{Karwowski and Pestka}(2007)}]{karw2}
\bibinfo{author}{\bibfnamefont{J.}~\bibnamefont{Karwowski}} \bibnamefont{and}
  \bibinfo{author}{\bibfnamefont{G.}~\bibnamefont{Pestka}},
  \bibinfo{journal}{Theoretical Chemistry Accounts}
  \textbf{\bibinfo{volume}{118}} (\bibinfo{year}{2007}),
  \urlprefix\url{https://link.springer.com/content/pdf/10.1007/s00214-007-0362-y.pdf}.

\bibitem[{\citenamefont{Villalba and Pino}(1998)}]{vill}
\bibinfo{author}{\bibfnamefont{V.~M.} \bibnamefont{Villalba}} \bibnamefont{and}
  \bibinfo{author}{\bibfnamefont{R.}~\bibnamefont{Pino}},
  \bibinfo{journal}{Physics Letters A} \textbf{\bibinfo{volume}{238}},
  \bibinfo{pages}{49} (\bibinfo{year}{1998}),
  \urlprefix\url{https://www.sciencedirect.com/science/article/pii/S0375960197008918}.

\bibitem[{\citenamefont{Karwowski}(2008)}]{karw}
\bibinfo{author}{\bibfnamefont{J.}~\bibnamefont{Karwowski}},
  \bibinfo{journal}{Journal of Physics: Conference Series}
  \textbf{\bibinfo{volume}{104}} (\bibinfo{year}{2008}),
  \urlprefix\url{https://iopscience.iop.org/article/10.1088/1742-6596/104/1/012033/pdf}.

\bibitem[{\citenamefont{Turbiner}(1988)}]{turb}
\bibinfo{author}{\bibfnamefont{A.}~\bibnamefont{Turbiner}},
  \bibinfo{journal}{Communications in Mathematical Physics}
  \textbf{\bibinfo{volume}{118}}, \bibinfo{pages}{467} (\bibinfo{year}{1988}),
  \urlprefix\url{https://projecteuclid.org/journals/communications-in-mathematical-physics/volume-118/issue-3/Quasi-exactly-solvable-problems-and-rm-sl2-algebra/cmp/1104162094.full}.

\bibitem[{\citenamefont{Liu and Hao}(2015)}]{liu}
\bibinfo{author}{\bibfnamefont{L.}~\bibnamefont{Liu}} \bibnamefont{and}
  \bibinfo{author}{\bibfnamefont{Q.}~\bibnamefont{Hao}},
  \bibinfo{journal}{Theoretical and Mathematical Physics}
  \textbf{\bibinfo{volume}{183}} (\bibinfo{year}{2015}),
  \urlprefix\url{https://link.springer.com/content/pdf/10.1007/s11232-015-0291-1.pdf}.

\bibitem[{\citenamefont{Downing and Portnoi}(2016)}]{down}
\bibinfo{author}{\bibfnamefont{C.~A.} \bibnamefont{Downing}} \bibnamefont{and}
  \bibinfo{author}{\bibfnamefont{M.~E.} \bibnamefont{Portnoi}},
  \bibinfo{journal}{Physical Review B} \textbf{\bibinfo{volume}{94}}
  (\bibinfo{year}{2016}),
  \urlprefix\url{https://journals.aps.org/prb/pdf/10.1103/PhysRevB.94.045430}.

\bibitem[{\citenamefont{Ho and Khalilov}(2000)}]{choo}
\bibinfo{author}{\bibfnamefont{C.-L.} \bibnamefont{Ho}} \bibnamefont{and}
  \bibinfo{author}{\bibfnamefont{V.~R.} \bibnamefont{Khalilov}},
  \bibinfo{journal}{Phys. Rev. A} \textbf{\bibinfo{volume}{61}}
  (\bibinfo{year}{2000}),
  \urlprefix\url{https://journals.aps.org/pra/pdf/10.1103/PhysRevA.61.032104}.

\bibitem[{\citenamefont{Chiang and Ho}(2002)}]{choo2}
\bibinfo{author}{\bibfnamefont{C.-M.} \bibnamefont{Chiang}} \bibnamefont{and}
  \bibinfo{author}{\bibfnamefont{C.-L.} \bibnamefont{Ho}},
  \bibinfo{journal}{Journal of Mathematical Physics}
  \textbf{\bibinfo{volume}{43}}, \bibinfo{pages}{43} (\bibinfo{year}{2002}),
  \urlprefix\url{https://aip.scitation.org/doi/abs/10.1063/1.1418426}.

\bibitem[{\citenamefont{Agboola}(2012)}]{agbo}
\bibinfo{author}{\bibfnamefont{Y.-Z.} \bibnamefont{Agboola},
  \bibfnamefont{Davis;~Zhang}}, \bibinfo{journal}{Modern physics letters A}
  \textbf{\bibinfo{volume}{27}} (\bibinfo{year}{2012}),
  \urlprefix\url{https://www.worldscientific.com/doi/pdf/10.1142\%2FS021773231250112X}.

\bibitem[{\citenamefont{Karwowski and Witek}(2016)}]{karw3}
\bibinfo{author}{\bibfnamefont{J.}~\bibnamefont{Karwowski}} \bibnamefont{and}
  \bibinfo{author}{\bibfnamefont{H.}~\bibnamefont{Witek}},
  \bibinfo{journal}{Molecular Physics} \textbf{\bibinfo{volume}{114}}
  (\bibinfo{year}{2016}),
  \urlprefix\url{https://www.tandfonline.com/doi/pdf/10.1080/00268976.2015.1115565}.

\bibitem[{\citenamefont{Akhmedov and Guseinova}(2009)}]{akhm}
\bibinfo{author}{\bibfnamefont{K.}~\bibnamefont{Akhmedov}} \bibnamefont{and}
  \bibinfo{author}{\bibfnamefont{N.}~\bibnamefont{Guseinova}},
  \bibinfo{journal}{Russian Physics Journal} \textbf{\bibinfo{volume}{52}},
  \bibinfo{pages}{321} (\bibinfo{year}{2009}),
  \urlprefix\url{https://link.springer.com/content/pdf/10.1007/s11182-009-9230-7.pdf}.

\bibitem[{\citenamefont{Pont et~al.}(2018)\citenamefont{Pont, Osenda, and
  Serra}}]{pont}
\bibinfo{author}{\bibfnamefont{F.~M.} \bibnamefont{Pont}},
  \bibinfo{author}{\bibfnamefont{O.}~\bibnamefont{Osenda}}, \bibnamefont{and}
  \bibinfo{author}{\bibfnamefont{P.}~\bibnamefont{Serra}},
  \bibinfo{journal}{Journal of Physics A: Mathematical and General}
  \textbf{\bibinfo{volume}{51}} (\bibinfo{year}{2018}),
  \urlprefix\url{https://iopscience.iop.org/article/10.1088/1751-8121/aab85e/pdf}.

\bibitem[{\citenamefont{Downing}(2017)}]{down2}
\bibinfo{author}{\bibfnamefont{C.~A.} \bibnamefont{Downing}},
  \bibinfo{journal}{Physical Review A} \textbf{\bibinfo{volume}{95}}
  (\bibinfo{year}{2017}),
  \urlprefix\url{https://journals.aps.org/pra/pdf/10.1103/PhysRevA.95.022105}.

\bibitem[{\citenamefont{Loos and Gill}(2012)}]{loos2}
\bibinfo{author}{\bibfnamefont{P.-F.} \bibnamefont{Loos}} \bibnamefont{and}
  \bibinfo{author}{\bibfnamefont{P.~M.~W.} \bibnamefont{Gill}},
  \bibinfo{journal}{Physical Review Letters} \textbf{\bibinfo{volume}{108}}
  (\bibinfo{year}{2012}),
  \urlprefix\url{https://journals.aps.org/prl/pdf/10.1103/PhysRevLett.108.083002}.

\bibitem[{\citenamefont{Guo et~al.}(2012)\citenamefont{Guo, REN, ZHOU, and
  GUO}}]{guo}
\bibinfo{author}{\bibfnamefont{G.-J.} \bibnamefont{Guo}},
  \bibinfo{author}{\bibfnamefont{Z.-Z.} \bibnamefont{REN}},
  \bibinfo{author}{\bibfnamefont{B.}~\bibnamefont{ZHOU}}, \bibnamefont{and}
  \bibinfo{author}{\bibfnamefont{X.-Y.} \bibnamefont{GUO}},
  \bibinfo{journal}{International Journal of Modern Physics B}
  \textbf{\bibinfo{volume}{26}} (\bibinfo{year}{2012}),
  \urlprefix\url{https://www.worldscientific.com/doi/pdf/10.1142\%2FS0217979212502013}.

\bibitem[{\citenamefont{Loos and Gill}(2009)}]{loos}
\bibinfo{author}{\bibfnamefont{P.-F.} \bibnamefont{Loos}} \bibnamefont{and}
  \bibinfo{author}{\bibfnamefont{P.~M.~W.} \bibnamefont{Gill}},
  \bibinfo{journal}{Physical Review Letters} \textbf{\bibinfo{volume}{103}}
  (\bibinfo{year}{2009}),
  \urlprefix\url{https://doi.org/10.1103/PhysRevLett.103.123008}.

\bibitem[{\citenamefont{Loos and Gill}(2010)}]{loos3}
\bibinfo{author}{\bibfnamefont{P.-F.} \bibnamefont{Loos}} \bibnamefont{and}
  \bibinfo{author}{\bibfnamefont{P.~M.~W.} \bibnamefont{Gill}},
  \bibinfo{journal}{Molecular Physics} \textbf{\bibinfo{volume}{108}}
  (\bibinfo{year}{2010}),
  \urlprefix\url{https://www.tandfonline.com/doi/pdf/10.1080/00268976.2010.508472}.

\bibitem[{\citenamefont{Jestädt et~al.}(2019)\citenamefont{Jestädt,
  Ruggenthaler, Oliveira, Rubio, and Appel}}]{QEDHAM}
\bibinfo{author}{\bibfnamefont{R.}~\bibnamefont{Jestädt}},
  \bibinfo{author}{\bibfnamefont{M.}~\bibnamefont{Ruggenthaler}},
  \bibinfo{author}{\bibfnamefont{M.~J.~T.} \bibnamefont{Oliveira}},
  \bibinfo{author}{\bibfnamefont{A.}~\bibnamefont{Rubio}}, \bibnamefont{and}
  \bibinfo{author}{\bibfnamefont{H.}~\bibnamefont{Appel}},
  \bibinfo{journal}{Advances in Physics} \textbf{\bibinfo{volume}{68}},
  \bibinfo{pages}{225} (\bibinfo{year}{2019}).

\bibitem[{\citenamefont{Frisk~Kockum et~al.}(2019)\citenamefont{Frisk~Kockum,
  Miranowicz, De~Liberato, Savasta, and Nori}}]{FriskKockum2019}
\bibinfo{author}{\bibfnamefont{A.}~\bibnamefont{Frisk~Kockum}},
  \bibinfo{author}{\bibfnamefont{A.}~\bibnamefont{Miranowicz}},
  \bibinfo{author}{\bibfnamefont{S.}~\bibnamefont{De~Liberato}},
  \bibinfo{author}{\bibfnamefont{S.}~\bibnamefont{Savasta}}, \bibnamefont{and}
  \bibinfo{author}{\bibfnamefont{F.}~\bibnamefont{Nori}},
  \bibinfo{journal}{Nature Reviews Physics} \textbf{\bibinfo{volume}{1}},
  \bibinfo{pages}{19} (\bibinfo{year}{2019}), ISSN \bibinfo{issn}{2522-5820},
  \urlprefix\url{https://doi.org/10.1038/s42254-018-0006-2}.

\bibitem[{\citenamefont{Sch\"afer et~al.}(2018)\citenamefont{Sch\"afer,
  Ruggenthaler, and Rubio}}]{PhysRevA.98.043801}
\bibinfo{author}{\bibfnamefont{C.}~\bibnamefont{Sch\"afer}},
  \bibinfo{author}{\bibfnamefont{M.}~\bibnamefont{Ruggenthaler}},
  \bibnamefont{and} \bibinfo{author}{\bibfnamefont{A.}~\bibnamefont{Rubio}},
  \bibinfo{journal}{Phys. Rev. A} \textbf{\bibinfo{volume}{98}},
  \bibinfo{pages}{043801} (\bibinfo{year}{2018}),
  \urlprefix\url{https://link.aps.org/doi/10.1103/PhysRevA.98.043801}.

\bibitem[{\citenamefont{Flick et~al.}(2018)\citenamefont{Flick, Schäfer,
  Ruggenthaler, Appel, and Rubio}}]{doi:10.1021/acsphotonics.7b01279}
\bibinfo{author}{\bibfnamefont{J.}~\bibnamefont{Flick}},
  \bibinfo{author}{\bibfnamefont{C.}~\bibnamefont{Schäfer}},
  \bibinfo{author}{\bibfnamefont{M.}~\bibnamefont{Ruggenthaler}},
  \bibinfo{author}{\bibfnamefont{H.}~\bibnamefont{Appel}}, \bibnamefont{and}
  \bibinfo{author}{\bibfnamefont{A.}~\bibnamefont{Rubio}},
  \bibinfo{journal}{ACS Photonics} \textbf{\bibinfo{volume}{5}},
  \bibinfo{pages}{992} (\bibinfo{year}{2018}).

\bibitem[{\citenamefont{Sidler et~al.}(2020)\citenamefont{Sidler, Ruggenthaler,
  Appel, and Rubio}}]{doi:10.1021/acs.jpclett.0c01556}
\bibinfo{author}{\bibfnamefont{D.}~\bibnamefont{Sidler}},
  \bibinfo{author}{\bibfnamefont{M.}~\bibnamefont{Ruggenthaler}},
  \bibinfo{author}{\bibfnamefont{H.}~\bibnamefont{Appel}}, \bibnamefont{and}
  \bibinfo{author}{\bibfnamefont{A.}~\bibnamefont{Rubio}},
  \bibinfo{journal}{The Journal of Physical Chemistry Letters}
  \textbf{\bibinfo{volume}{11}}, \bibinfo{pages}{7525} (\bibinfo{year}{2020}),
  \bibinfo{note}{pMID: 32805122},
  \urlprefix\url{https://doi.org/10.1021/acs.jpclett.0c01556}.

\bibitem[{\citenamefont{Lacombe et~al.}(2019)\citenamefont{Lacombe, Hoffmann,
  and Maitra}}]{PhysRevLett.123.083201}
\bibinfo{author}{\bibfnamefont{L.}~\bibnamefont{Lacombe}},
  \bibinfo{author}{\bibfnamefont{N.~M.} \bibnamefont{Hoffmann}},
  \bibnamefont{and} \bibinfo{author}{\bibfnamefont{N.~T.}
  \bibnamefont{Maitra}}, \bibinfo{journal}{Phys. Rev. Lett.}
  \textbf{\bibinfo{volume}{123}}, \bibinfo{pages}{083201}
  (\bibinfo{year}{2019}),
  \urlprefix\url{https://link.aps.org/doi/10.1103/PhysRevLett.123.083201}.

\bibitem[{\citenamefont{Andolina et~al.}(2019)\citenamefont{Andolina,
  Pellegrino, Giovannetti, MacDonald, and Polini}}]{PhysRevB.100.121109}
\bibinfo{author}{\bibfnamefont{G.~M.} \bibnamefont{Andolina}},
  \bibinfo{author}{\bibfnamefont{F.~M.~D.} \bibnamefont{Pellegrino}},
  \bibinfo{author}{\bibfnamefont{V.}~\bibnamefont{Giovannetti}},
  \bibinfo{author}{\bibfnamefont{A.~H.} \bibnamefont{MacDonald}},
  \bibnamefont{and} \bibinfo{author}{\bibfnamefont{M.}~\bibnamefont{Polini}},
  \bibinfo{journal}{Phys. Rev. B} \textbf{\bibinfo{volume}{100}},
  \bibinfo{pages}{121109} (\bibinfo{year}{2019}),
  \urlprefix\url{https://link.aps.org/doi/10.1103/PhysRevB.100.121109}.

\bibitem[{\citenamefont{Schuler et~al.}(2020)\citenamefont{Schuler, Bernardis,
  Läuchli, and Rabl}}]{10.21468/SciPostPhys.9.5.066}
\bibinfo{author}{\bibfnamefont{M.}~\bibnamefont{Schuler}},
  \bibinfo{author}{\bibfnamefont{D.~D.} \bibnamefont{Bernardis}},
  \bibinfo{author}{\bibfnamefont{A.~M.} \bibnamefont{Läuchli}},
  \bibnamefont{and} \bibinfo{author}{\bibfnamefont{P.}~\bibnamefont{Rabl}},
  \bibinfo{journal}{SciPost Phys.} \textbf{\bibinfo{volume}{9}},
  \bibinfo{pages}{66} (\bibinfo{year}{2020}),
  \urlprefix\url{https://scipost.org/10.21468/SciPostPhys.9.5.066}.

\bibitem[{\citenamefont{Settineri et~al.}(2021)\citenamefont{Settineri,
  Di~Stefano, Zueco, Hughes, Savasta, and Nori}}]{PhysRevResearch.3.023079}
\bibinfo{author}{\bibfnamefont{A.}~\bibnamefont{Settineri}},
  \bibinfo{author}{\bibfnamefont{O.}~\bibnamefont{Di~Stefano}},
  \bibinfo{author}{\bibfnamefont{D.}~\bibnamefont{Zueco}},
  \bibinfo{author}{\bibfnamefont{S.}~\bibnamefont{Hughes}},
  \bibinfo{author}{\bibfnamefont{S.}~\bibnamefont{Savasta}}, \bibnamefont{and}
  \bibinfo{author}{\bibfnamefont{F.}~\bibnamefont{Nori}},
  \bibinfo{journal}{Phys. Rev. Research} \textbf{\bibinfo{volume}{3}},
  \bibinfo{pages}{023079} (\bibinfo{year}{2021}),
  \urlprefix\url{https://link.aps.org/doi/10.1103/PhysRevResearch.3.023079}.

\bibitem[{\citenamefont{Rokaj et~al.}(2021)\citenamefont{Rokaj, Ruggenthaler,
  Eich, and Rubio}}]{rokaj2021free}
\bibinfo{author}{\bibfnamefont{V.}~\bibnamefont{Rokaj}},
  \bibinfo{author}{\bibfnamefont{M.}~\bibnamefont{Ruggenthaler}},
  \bibinfo{author}{\bibfnamefont{F.~G.} \bibnamefont{Eich}}, \bibnamefont{and}
  \bibinfo{author}{\bibfnamefont{A.}~\bibnamefont{Rubio}},
  \emph{\bibinfo{title}{The free electron gas in cavity quantum
  electrodynamics}} (\bibinfo{year}{2021}), \eprint{2006.09236}.

\bibitem[{\citenamefont{Shin and Metiu}(1995)}]{doi:10.1063/1.468795}
\bibinfo{author}{\bibfnamefont{S.}~\bibnamefont{Shin}} \bibnamefont{and}
  \bibinfo{author}{\bibfnamefont{H.}~\bibnamefont{Metiu}},
  \bibinfo{journal}{The Journal of Chemical Physics}
  \textbf{\bibinfo{volume}{102}}, \bibinfo{pages}{9285} (\bibinfo{year}{1995}),
  \urlprefix\url{https://doi.org/10.1063/1.468795}.

\bibitem[{\citenamefont{Mandal et~al.}(2020{\natexlab{b}})\citenamefont{Mandal,
  Krauss, and Huo}}]{acs.jpcb.0c03227}
\bibinfo{author}{\bibfnamefont{A.}~\bibnamefont{Mandal}},
  \bibinfo{author}{\bibfnamefont{T.~D.} \bibnamefont{Krauss}},
  \bibnamefont{and} \bibinfo{author}{\bibfnamefont{P.}~\bibnamefont{Huo}},
  \bibinfo{journal}{The Journal of Physical Chemistry B}
  \textbf{\bibinfo{volume}{124}}, \bibinfo{pages}{6321}
  (\bibinfo{year}{2020}{\natexlab{b}}), \bibinfo{note}{pMID: 32589846}.

\bibitem[{\citenamefont{Power et~al.}(1959)\citenamefont{Power, Zienau, and
  Massey}}]{Zienau}
\bibinfo{author}{\bibfnamefont{E.~A.} \bibnamefont{Power}},
  \bibinfo{author}{\bibfnamefont{S.}~\bibnamefont{Zienau}}, \bibnamefont{and}
  \bibinfo{author}{\bibfnamefont{H.~S.~W.} \bibnamefont{Massey}},
  \bibinfo{journal}{Philosophical Transactions of the Royal Society of London.
  Series A, Mathematical and Physical Sciences} \textbf{\bibinfo{volume}{251}},
  \bibinfo{pages}{427} (\bibinfo{year}{1959}),
  \eprint{https://royalsocietypublishing.org/doi/pdf/10.1098/rsta.1959.0008},
  \urlprefix\url{https://royalsocietypublishing.org/doi/abs/10.1098/rsta.1959.0008}.

\bibitem[{\citenamefont{Ruggenthaler et~al.}(2014)\citenamefont{Ruggenthaler,
  Flick, Pellegrini, Appel, Tokatly, and Rubio}}]{PhysRevA.90.012508}
\bibinfo{author}{\bibfnamefont{M.}~\bibnamefont{Ruggenthaler}},
  \bibinfo{author}{\bibfnamefont{J.}~\bibnamefont{Flick}},
  \bibinfo{author}{\bibfnamefont{C.}~\bibnamefont{Pellegrini}},
  \bibinfo{author}{\bibfnamefont{H.}~\bibnamefont{Appel}},
  \bibinfo{author}{\bibfnamefont{I.~V.} \bibnamefont{Tokatly}},
  \bibnamefont{and} \bibinfo{author}{\bibfnamefont{A.}~\bibnamefont{Rubio}},
  \bibinfo{journal}{Phys. Rev. A} \textbf{\bibinfo{volume}{90}},
  \bibinfo{pages}{012508} (\bibinfo{year}{2014}),
  \urlprefix\url{https://link.aps.org/doi/10.1103/PhysRevA.90.012508}.

\bibitem[{Sup(2021)}]{Supp}
 (\bibinfo{year}{2021}), \bibinfo{note}{see Supplemental Material at}.

\end{thebibliography}
\end{document}